\documentclass[review]{elsarticle}
\makeatletter
\def\ps@pprintTitle{%
 \let\@oddhead\@empty
 \let\@evenhead\@empty
 \def\@oddfoot{\centerline{\thepage}}%
 \let\@evenfoot\@oddfoot}
\makeatother

\pdfoutput=1
\usepackage{graphicx}
\usepackage{subfigure}
\usepackage{amsfonts, amsmath, amssymb}
\usepackage{mathabx}
\usepackage{booktabs,siunitx}
\usepackage[svgnames,table]{xcolor}
\usepackage[tableposition=above]{caption}
\usepackage{pifont}
\usepackage{bm}
\usepackage{cancel}
\usepackage{breqn}
\usepackage{caption}
\usepackage{color}
\usepackage{enumerate}
\usepackage{float}
\usepackage{hyperref}
\usepackage{soul}
\usepackage[english]{babel}
\usepackage[margin=1in]{geometry}
\usepackage{mathtools}
\usepackage{multirow}
\usepackage{setspace}
\usepackage{subfigure}
\usepackage{stmaryrd}
\usepackage{lineno}
\DeclareUnicodeCharacter{00B3}{\textsuperscript{3}}
\usepackage[ruled,linesnumbered]{algorithm2e}
\RestyleAlgo{ruled}
\SetKwComment{Comment}{/* }{ */}

\SetCommentSty{mycommfont}
\DeclareMathAlphabet{\mathpzc}{OT1}{pzc}{m}{it}

\newcommand \dd[2] {\frac{{\rm d} #1}{{\rm d} #2}}
\newcommand \dD[2] {\frac{{\rm d^2} #1}{{\rm d} {#2}^2}}
\renewcommand \d[1]{{\rm{d}} #1}
\newcommand \D [2]{\frac{\partial #1}{\partial #2}}
\newcommand \DD[2]{\frac{\partial^2 #1}{\partial #2 ^2}}

\renewcommand{\vec}[1]{\bm{\mathrm{#1}}}
\newcommand{\V}[1]{\bm{\mathrm{#1}}}

\def \grad{\nabla}

\def \x{\vec{x}}

\def \q{\vec{q}}
\def \n{\vec{n}}
\def \s{\vec{s}}
\def \u{\vec{u}}

\def \grad{\nabla}

\def \rhos{\rho^{\rm S}}
\def \rhol{\rho^{\rm L}}

\def \rhov{\rho^{\rm V}}
\def \rrhols{R_\rho^{\rm LS}}
\def \rrhosl{R_\rho^{\rm SL}}
\def \rrhovl{R_\rho^{\rm VL}}
\def \rrholv{R_\rho^{\rm LV}}
\def \rrhovs{R_\rho^{\rm VS}}
\def \cps{C^{\rm S}}
\def \cpl{C^{\rm L}}

\def \cpv{C^{\rm V}}
\def \Ts{T^{\rm S}}
\def \Tl{T^{\rm L}}

\def \Tv{T^{\rm V}}
\def \ks{\kappa^{\rm S}}
\def \kl{\kappa^{\rm L}}

\def \kv{\kappa^{\rm V}}
\def \hs{h^{\rm S}}
\def \hl{h^{\rm L}}
\def \hv{h^{\rm V}}
\def \ul{u^{\rm L}}
\def \us{u^{\rm S}}

\def \uv{u^{\rm V}}
\def \uln{u^{\rm L}_{\rm n}}
\def \usn{u^{\rm S}_{\rm n}}

\def \uvn{u^{\rm V}_{\rm n}}
\def \el{e^{\rm L}}
\def \es{e^{\rm S}}
\def \ev{e^{\rm V}}

\def \pl{p^{\rm L}}
\def \ps{p^{\rm S}}
\def \pv{p^{\rm V}}
\def \vul{\vec{u}^{\rm L}}
\def \vuv{\vec{u}^{\rm V}}
\def \vus{\vec{u}^{\rm S}}

\def \Omegas{\Omega^{\rm S}}
\def \Omegal{\Omega^{\rm L}}

\def \Omegav{\Omega^{\rm V}}

\def \alphas{\alpha^{\rm S}}
\def \alphal{\alpha^{\rm L}}
\def \alphav{\alpha^{\rm V}}

\def \pl{p^{\rm L}}
\def \ps{p^{\rm S}}
\def \etal{\eta^{\rm L}}
\def \etas{\eta^{\rm S}}
\def \etav{\eta^{\rm V}}

\def \half{\frac{1}{2}}
\def \threehalf{\frac{3}{2}}

\graphicspath{{figures/}}

\usepackage{caption}  

\newcommand{\upperRomannumeral}[1]{\uppercase\expandafter{\romannumeral#1}}

\newcommand{\REVIEW}[1]{{#1}}

\newcommand{\bluecancel}[1]{\renewcommand\CancelColor{\color{blue}}\cancel{#1}}
\newcommand{\redcancel}[1]{\renewcommand\CancelColor{\color{red}}\cancel{#1}}
\newcommand{\blackcancel}[1]{\renewcommand\CancelColor{\color{black}}\cancel{#1}}

\begin{document}
\let\today\relax
\let\underbrace\LaTeXunderbrace
\let\overbrace\LaTeXoverbrace

\begin{frontmatter}

\title{Analytical and numerical solutions to the three-phase Stefan problem with simultaneous occurrences of melting, solidification, boiling, and condensation phenomena}
\author[SDSU]{Mehran Soleimani}
\author[CSM]{Kimmo Koponen}
\author[CSM]{Nils Tilton}
\author[SDSU]{Amneet Pal Singh Bhalla\corref{mycorrespondingauthor}}
\ead{asbhalla@sdsu.edu}

\address[SDSU]{Department of Mechanical Engineering, San Diego State University, San Diego, CA}
\address[CSM]{Department of Mechanical Engineering, Colorado School of Mines, Golden, CO}
\cortext[mycorrespondingauthor]{Corresponding author}

\begin{abstract}
The one-dimensional (1D) Stefan problem is a prototypical heat and mass transfer problem that analyzes the temperature distribution in a material undergoing phase change. In addition, it describes the evolution of the phase change front within the phase change material (PCM). Analytical solutions to the two-phase Stefan problem that describe melting  of a solid  or boiling  of a liquid have been extensively discussed in the literature. Density change effects and associated fluid flow phenomena during phase change are typically ignored to simplify the analysis. As the PCM boils or condenses, it undergoes a density change of 1000 or more. The effects of density changes and convection cannot be ignored when dealing with such problems. In our recent work~\cite{thirumalaisamy2023low}, we found analytical solutions to the two-phase Stefan problem that account for a jump in the thermophysical properties of the two phases, including density. In the present work, we extend our prior analyses to obtain analytical solutions to the three-phase Stefan problem in which an initially solid PCM melts and boils under imposed temperature conditions. This scenario is typical of metal additive manufacturing (AM) and welding processes, wherein a high-power laser melts and boils the metal powder or substrate. Each phase of the PCM (solid, liquid, and vapor) has distinct thermophysical properties, including density, thermal conductivity, and heat capacity. While deriving the analytical solution, all relevant jump conditions, including density and kinetic energy, are accounted for. It is shown that the three-phase Stefan problem admits similarity transformations and similarity solutions. Similarity solutions to the three-phase Stefan problem require solving two coupled transcendental equations. To our knowledge, this is the first work that presents an analytical solution to the three-phase Stefan problem with simultaneous \underline{m}elting, \underline{s}olidificatio\underline{n}, \underline{b}oiling, and \underline{c}ondensation (MSNBC). Furthermore, we describe a numerical method for solving the three-phase Stefan problem with second-order accuracy. Numerical solutions for vapor-liquid and liquid-solid interface positions and temperature distributions are compared to analytical solutions. It is demonstrated that the proposed sharp-interface technique can simulate phase change phenomena in moving domains with second-order spatiotemporal accuracy.           
\end{abstract}


\end{frontmatter}

\section{Introduction}\label{introduction}

Stefan's problem describes the evolution of the boundary between two phases of a material undergoing a phase change, such as ice melting into water. Josef Stefan proposed the Stefan problem in a set of four papers in 1890. Stefan's work built on a similar problem encountered in 1831 by Lam\'e and Clapeyron while studying the solidification of the earth's crust.  A mathematical solution to Lam\'e and Clapeyron's problem was subsequently provided by Neumann in 1860. Stefan consolidated the work of Lam\'e, Clapeyron, Neumann and others into precise mathematical statements involving free-moving boundaries, which now bear his name~\cite{vuik1993some,rubinvsteuin2000stefan}. 

Most of the prior literature has analyzed Stefan's problems analytically by ignoring density variations in the two phases~\cite{hahn2012heat}. Whenever density changes during phase change, fluid flow is induced. This makes the heat transfer problem more complex since it is coupled with fluid flow. Alexiades and Solomon's textbook~\cite{alexiades2018mathematical} provides an analytical solution to the two-phase Stefan problem with density difference between the two phases despite its apparent difficulty. Surprisingly, the analytical solution that includes a density jump is quite similar to the solution that ignores the density variation. Unfortunately, the textbook solution to the two-phase Stefan problem by Alexiades and Solomon is not well known. The standard two-phase Stefan problem, in which only the heat equation is used and density change-induced fluid flow is not involved, remains the gold standard for validating advanced computational fluid dynamics (CFD) algorithms for modeling melting and solidification~\cite{huang2022consistent,yan2018fully,javierre2006comparison}, as well as boiling and condensation~\cite{gibou2007level,khalloufi2020adaptive} phenomena. Alexiades and Solomon drop the kinetic energy jump term from the Stefan condition because they believed that it ``destroys" the similarity solution. However, we recently demonstrated that this is not the case, and demonstrated how to incorporate the kinetic energy jump into the two-phase Stefan problem while retaining the similarity solution. As part of our previous works~\cite{thirumalaisamy2023low,thirumalaisamy2025consistent}, we considered a number of test problems involving melting and solidification of a (metallic) PCM to demonstrate the effects of density changes on volume shrinkage or expansion of the PCM during phase change. 

The current work extends our previous work on two-phase Stefan problems to three-phase  problems in which melting/solidification of the PCM is accompanied by its boiling/condensation. This work is motivated by our ongoing efforts to develop fully-resolved CFD algorithms for modeling metal AM processes in which a metallic powder or substrate is subjected to MSNBC through a high-power laser. Currently, state-of-the-art CFD solvers only model melting and solidification of metallic PCMs in a fully-resolved manner, adding (ad-hoc) phenomenological models to account for evaporation and condensation phenomena~\cite{yan2018fully,yu2022quantitative,ai2017three,ai2017prediction}. In the discretized equations, this requires the introduction of artificial terms, such as the ``recoil pressure." Because no analytical solution to the three-phase Stefan problem exists in the literature to date, the results of CFD simulations involving MSNBC have not been validated in a principled manner. The current work fills this essential gap.  

In Sec.~\ref{sec_3p_derivation} of this work, we derive the analytical solution to the three-phase Stefan problem. The imposed temperature boundary condition causes an initially solid PCM to melt and boil. Each phase has distinct thermophysical properties, including density, and we account for fluid flow and convective heat transfer effects. The analytical solutions are based on similarity transformations of the heat equations describing the evolution of temperature in the solid, liquid, and vapor phases of the PCM. For the convenience of the reader, all steps of the derivation are provided in detail. Sec.~\ref{sec_sharp_interface_method} presents a sharp-interface method for solving the one-dimensional three-phase Stefan problem with MSNBC, achieving second-order spatial and temporal accuracy. This is demonstrated  by comparing the numerical and analytical solutions for two- and three-phase Stefan problems.

\section{Analytical solution to the three-phase Stefan problem}\label{sec_3p_derivation}

\begin{figure}[]
	\begin{center}
       \includegraphics[scale=0.4]{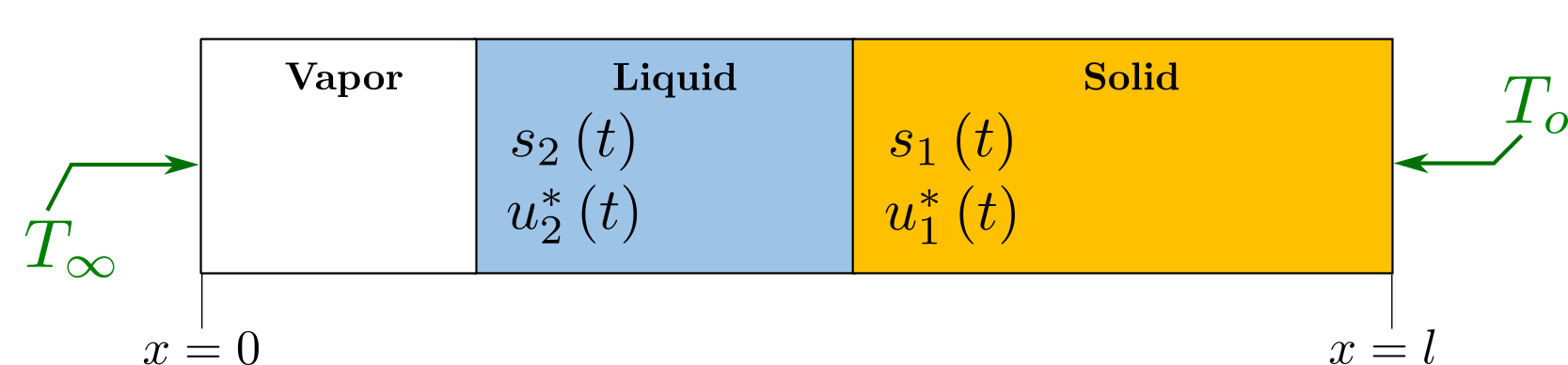}
	\end{center}
\caption{Schematic of the three-phase Stefan problem in which an initially solid PCM melts and boils due to an imposed temperature condition at $x = 0$.}\label{fig_3pstefan_schematic}
\end{figure}

We consider a phase change material (PCM) with a melting temperature $T_m$ and a vaporization temperature $T_v$. The PCM occupies the 1D domain $\Omega := 0 \le x \le l$. At the initial time $t = 0$, the full domain is occupied by the solid phase, with a uniform initial temperature $T_i < T_m$. The right boundary ($x = l$) is open, while the left boundary ($x= 0$) is closed. Imposing a temperature of $T_\infty > T_v$ at the left boundary melts and boils the PCM. The thermophysical properties of the solid phase are $\rhos$, $\cps$, and $\ks$, those of the liquid phase are $\rhol$, $\cpl$, and $\kl$, and those of the vapor phase are $\rhov$, $\cpv$, and $\kv$. In each phase, these properties are assumed to be constant\footnote{\REVIEW{This assumption is made to solve the Stefan problem analytically. In reality, thermophysical properties of a phase depend on its temperature.}}. When the PCM melts, the melt front $x = s_1(t)$ moves rightward. Due to the imposed temperature $T_\infty > T_v$, the molten PCM begins to boil and forms a boiling front $x = s_2(t)$; see Fig.~\ref{fig_3pstefan_schematic}. At $t = 0^{+}$, both fronts appear simultaneously at $x =0$, but due to the realistic thermophysical properties of the PCM, the melt front moves faster than the boiling front. \REVIEW{\ref{sec_compare_interface_speed} provides a ``back-of-the-envelope" comparison of the melt and boiling front speeds}. The solid-to-liquid density ratio is denoted $\rrhosl = \rhos/\rhol$, whereas its inverse is denoted $\rrhols = \rhol/\rhos$. Similarly, the liquid-to-vapor and vapor-to-liquid density ratios are denoted $\rrholv = \rhol/\rhov$ and $\rrhovl = \rhov/\rhol$, respectively. 

The conservation of mass, momentum and energy equations in the solid, liquid, and vapor phases read as
\begin{subequations} \label{eqn_nstd_sp}
\begin{alignat}{2}
&\D{\rho}{t} + \V{\grad} \cdot (\rho \u ) = 0 ,\label{eqn_mass_cons} \\ 
&\D{\rho \u}{t} + \V{\grad} \cdot (\rho \u \otimes \u + p \mathbb{I}) -  \sigma \mathcal{C} \delta (\x - \s) \n = 0,  \label{eqn_mom_cons} \\
&\D{}{t}\left(\rho \left[e + \frac{1}{2}| \u|^2 \right] \right) + \V{\grad} \cdot \left( \rho\left[ e + \frac{1}{2} | \u |^2 \right] \u + \q + p \u \right) - \sigma \mathcal{C} \delta (\x - \s) \u \cdot \n = 0. 
 \label{eqn_energy_cons}
\end{alignat}
\end{subequations}
In the momentum Eq.~\eqref{eqn_mom_cons}, $\sigma$ is the surface tension coefficient between the two phases (vapor-liquid and liquid-solid in this context), $\mathcal{C}$ represents the mean local curvature of the interface, $\s$ represents the position of the interface, $\delta$ is the Dirac delta distribution, and $\n$ is the outward unit normal vector of the interface (pointing in the positive $x$ direction in Fig.~\ref{fig_3pstefan_schematic}). In the energy  Eq.~\eqref{eqn_energy_cons}, $e$ denotes the internal energy and $\q = -\kappa \V{\grad} T$ is the conductive heat flux. The melt front moves with a velocity $u_1^* = \d{s_1}/\d{t}$ and the boiling front moves with a velocity $u_2^* = \d{s_2}/\d{t}$. We ignore the viscous stress tensor in the momentum Eq.~\eqref{eqn_mom_cons} as shear forces are zero in a one-dimensional setting. 

\subsection{Jump conditions at the boiling and melt fronts}

Applying the Rankine–Hugoniot condition for the mass balance Eq.~\eqref{eqn_mass_cons} across the vapor-liquid interface gives 
\begin{equation}\label{eqn_mass_jump_V-L}
	\left( \rhov - \rhol \right) u^*_2 = (\rhov \vuv - \rhol \vul) \cdot \n =  \rhov \uvn - \rhol \uln.
\end{equation}	
Here, $u_2^* = \d{s_2}/\d{t} = \u \cdot \n$ represents the normal velocity of the vapor-liquid interface, $\vul$ and $\vuv$ are velocities of the liquid and vapor phases, respectively, and $\uvn = \vuv \cdot \n$ and $\uln = \vul \cdot \n$ are the vapor and liquid velocities in the immediate neighborhood of the boiling front. The constant density in each phase indicates that the velocity in each phase is spatially uniform; see the mass balance Eq.~\eqref{eqn_mass_cons}. Thus, $\uvn$ and $\uln$ are spatially uniform vapor and liquid velocities. The velocity in the vapor domain $\Omegav(t) := 0 \le x < s_2(t)$ is zero, i.e., $\uv \equiv 0$. This follows from the 1D mass balance Eq.~\eqref{eqn_mass_cons} and the zero-velocity (closed boundary) condition at the left boundary $x = 0$. Substituting $\uvn = 0$ in Eq.~\eqref{eqn_mass_jump_V-L}, we get the liquid domain velocity 
\begin{equation}\label{eqn_uln}
		\uln = \left( 1 - \frac{\rhov}{\rhol} \right) u^*_2 = \left(1 - \rrhovl\right)u^*_2.		
\end{equation}	
Similarly, applying the mass jump condition across the liquid-solid interface gives 
\begin{equation}
    \left(\rhol - \rhos \right) u^*_1 =  (\rhol\vul - \rhos\vus)\cdot \n = \rhol \uln - \rhos \usn.
		\label{eqn_mass_jump_S-L}		
\end{equation}
Here, $\vus$ denotes velocity in the solid phase. Substituting Eq.~\eqref{eqn_uln} into Eq.~\eqref{eqn_mass_jump_S-L}, the solid domain velocity can be determined
\begin{equation}\label{eqn_usn}
\usn =  \left( \frac{\rhol}{\rhos} - \frac{\rhov}{\rhos} \right) u^*_2 + \left( 1 - \frac{\rhol}{\rhos} \right)u^*_1 = \left(\rrhols - \rrhovs \right)u^*_2 + \left(1- \rrhols \right)u^*_1.
\end{equation}
After determining the solid and liquid domain velocities, we compute $ u^*_1 - \usn $ and $\usn - \uln$, which will be utilized in the subsequent steps of the derivation,
\begin{subequations} \label{eqn_u1-us-ul}
\begin{alignat}{2}
u^*_1 - \usn &= \frac{\rhol}{\rhos}u^*_1- \left(\frac{\rhol}{\rhos} - \frac{\rhov}{\rhos} \right) u^*_2, \label{eqn_u1-us} \\
\usn - \uln &= \left( \frac{\rhol}{\rhos} - \frac{\rhov}{\rhos} \right) u^*_2 + \left( 1 - \frac{\rhol}{\rhos} \right)u^*_1- \left( 1 - \frac{\rhov}{\rhol} \right) u^*_2. \label{eqn_us-ul}
\end{alignat}
\end{subequations}

Next, applying the Rankine–Hugoniot condition for the energy conservation Eq.~\eqref{eqn_energy_cons} across the vapor-liquid interface yields the Stefan condition of interface 2 
\begin{equation}\label{eqn_energy_jump_VL}
\rhov \left( \ev + \frac{(\uvn)^2}{2} \right) \left(u^*_2 - \uvn\right) - \rhol \left( \el + \frac{(\uln)^2}{2} \right) (u^*_2 - \uln) 
+ \left(\pl \uln - \pv \uvn\right)  - \sigma_2 \mathcal{C}_2 u^*_2 = \left[{\q}^{\rm V} - {\q}^{\rm L}  \right] \cdot \n.
\end{equation} 
By setting $\uv=0$ in the the equation above, we get a simplified relation:
\begin{equation}\label{eqn_simplified_energy_jump_VL}
\rhov \ev u^*_2 - \rhol \left(\el + \frac{(\uln)^2}{2}\right) \left(u^*_2 - \uln \right) + \pl \uln - \sigma_2 \mathcal{C}_2 u^*_2 
		= \left[{\q}^{\rm V} - {\q}^{\rm L}  \right] \cdot \n.
\end{equation}
Substituting $ u^*_2 - \uln$ from Eq.~\eqref{eqn_uln} into Eq.~\eqref{eqn_simplified_energy_jump_VL} yields	\begin{equation}\label{eqn_simplified2_energy_jump_VL}
\rhov \ev u^*_2 - \rhov \left(\el + \frac{(\uln)^2}{2}\right) u^*_2 + \pl \uln - \sigma_2 \mathcal{C}_2 u^*_2  = \left[{\q}^{\rm V} - {\q}^{\rm L}  \right] \cdot \n.
\end{equation}
When $\uln$ from Eq.~\eqref{eqn_uln} is substituted into the above equation, it gives us	\begin{equation}\label{eqn_simplified2_energy_jump_VL}
\rhov \left(\ev- \el - \frac{(\uln)^2}{2}\right) u^*_2 + \pl\left( 1 - \frac{\rhov}{\rhol} \right) u^*_2  - \sigma_2 \mathcal{C}_2 u^*_2 
		= \left[{\q}^{\rm V} - {\q}^{\rm L}  \right] \cdot \n.
\end{equation}

\noindent Enthalpy in each phase can be expressed in terms of its internal energy $e$ as follows,
\begin{subequations} \label{eqn_enthalpies}
\begin{alignat}{2}
&\hs = \cps \left(\Ts - T_r\right) = \es + \frac{\ps}{\rhos}, \label{eqn_enthalpy_S} \\
&\hl = \cpl \left(\Tl - T_r\right) + L_m = \el + \frac{\pl}{\rhol}, \label{eqn_enthalpy_L} \\  
&\hv = \cpv \left(\Tv - T_r\right) + L_m + L_e = \ev + \frac{\pv}{\rhov}. \label{eqn_enthalpy_V}
\end{alignat}
\end{subequations}
Here, $L_m$ and $L_e$ denote the latent heat of melting and vaporization, respectively, $p^\Gamma(x,t)$ and $T^\Gamma(x,t)$ are the thermodynamic pressure and temperature of phase $\Gamma$ = \{S,L,V\}, and $T_r$ is the reference temperature for measuring enthalpies.

Expressing $\ev$ and $\el$ in terms of temperature (see Eqs.~\eqref{eqn_enthalpy_V} and~\eqref{eqn_enthalpy_L}), the boiling front's Stefan condition (Eq.~\eqref{eqn_simplified2_energy_jump_VL}) becomes
\begin{align}\label{eqn_simplified3_energy_jump_VL}
&\rhov \left[ \left(\cpv - \cpl\right)\left(T_v - T_r\right) + L_e -  \frac{\pv}{\rhov} +  \frac{\pl}{\rhol} - \frac{(\uln)^2}{2} \right] u^*_2 + \pl\left( 1 - \frac{\rhov}{\rhol} \right) u^*_2  - \sigma_2 \mathcal{C}_2 u^*_2  = \left[{\q}^{\rm V} - {\q}^{\rm L}  \right] \cdot \n, \\
\hookrightarrow & \rhov\left[  \left(\cpv - \cpl\right)\left(T_v - T_r\right) + L_e - \frac{(\uln)^2}{2} \right] u^*_2  + \left( -\pv + \cancel{\frac{\rhov}{\rhol}\pl} + \pl -\cancel{\frac{\rhov}{\rhol}\pl} - \sigma_2 \mathcal{C}_2 \right) u^*_2   = \left[{\q}^{\rm V} - {\q}^{\rm L}  \right] \cdot \n.
\end{align}
Here, we have used the saturation temperature condition $\Tv(s_2,t) = \Tl(s_2,t) = T_v$ at the boiling front.

To further simply Eq.~\eqref{eqn_simplified3_energy_jump_VL}, we require the pressure jump $ \pl-\pv$ across the vapor-liquid interface. This is obtained by invoking the Rankine–Hugoniot condition on the momentum  Eq.~\eqref{eqn_mom_cons} across the vapor-liquid interface, 
\begin{equation}
\label{eqn_momentum1_jump_VL}
\left(\rhov \uvn - \rhol \uln \right) u^*_2 =
		 \rhov \left(\uvn\right)^2 - \rhol \left(\uln\right)^2 + \pv- \pl + \sigma_2 \mathcal{C}_2. 
\end{equation}
Setting $ \uvn=0$ and using Eq.~\eqref{eqn_uln}, the above equation can be written as
\begin{align}
\pl- \pv &= \rhol \uln (u^*_2 -  \uln) + \sigma_2 \mathcal{C}_2 = \rhov \uln u^*_2 + \sigma_2 \mathcal{C}_2. \label{eqn_momentum2_jump_VL}
\end{align}

\noindent In light of Eqs.~\eqref{eqn_momentum2_jump_VL} and~\eqref{eqn_uln}, the Stefan condition at the vapor-liquid interface becomes
\begin{align}
&\rhov \left[  \left(\cpv - \cpl\right)\left(T_v - T_r\right) + L_e - \frac{(\uln)^2}{2} \right] u^*_2 + \left(\rhov \uln u^*_2 + \cancel{\sigma_2 \mathcal{C}_2}  -\cancel{\sigma_2 \mathcal{C}_2} \right) u^*_2  = \left[{\q}^{\rm V} - {\q}^{\rm L}  \right] \cdot \n, \nonumber \\
\hookrightarrow & \rhov \left[  \left(\cpv - \cpl\right)\left(T_v - T_r\right) + L_e   + \underbrace{\frac{1}{2} \left(1 - \left(\frac{\rhov}{\rhol}\right)^2  \right) (u^*_2)^2}_{\textrm{kinetic energy}} \right] u^*_2  = \left[{\q}^{\rm V} - {\q}^{\rm L}  \right] \cdot \n. \label{eqn_simplified4_energy_jump_VL}
\end{align}

\noindent Next, we consider the jump conditions at the liquid-solid melt front located at $x = s_1(t)$.  The Rankine-Hugonoit condition for the mass balance equation yields 
\begin{subequations} \label{eqn_mass_jump_LS}
\begin{alignat}{2}
&\left(\rhol - \rhos \right) u^*_1 =  \rhol \uln - \rhos \usn, \label{eqn_mass_jump_LSa} \\
\hookrightarrow & \rhol(u^*_1 -\uln) = \rhos(u^*_1 -\usn). \label{eqn_mass_jump_LSb}
\end{alignat}
\end{subequations}
Similarly, the energy jump condition across the liquid-solid interface reads as
\begin{equation}\label{eqn_energy_jump_LS}
\rhol \left( \el + \frac{(\uln)^2}{2} \right) (u^*_1 - \uln) - \rhos \left( \es + \frac{(\usn)^2}{2} \right) (u^*_1 - \usn) + \left(\ps \usn - \pl \uln\right)  - \sigma_1 \mathcal{C}_1 u^*_1 = \left[{\q}^{\rm L} - {\q}^{\rm S}  \right] \cdot \n. 
\end{equation} 

\noindent Substituting Eq.~\eqref{eqn_mass_jump_LS} into Eq.~\eqref{eqn_energy_jump_LS} simplifies the above equation to 	
\begin{equation}\label{eqn_simplified2_energy_jump_LS}
\rhos \left(\el + \frac{(\uln)^2}{2} - \es - \frac{(\usn)^2}{2} \right) \left(u^*_1 - \usn \right) + \ps \usn - \pl \uln - \sigma_1 \mathcal{C}_1 u^*_1 
= \left[{\q}^{\rm L} - {\q}^{\rm S}  \right] \cdot \n.
\end{equation}

\noindent Expressing $ \el $ and $\es$ in terms of enthalpy (see Eqs.~\eqref{eqn_enthalpies}) and using the saturation temperature condition $\Tl(s_1,t) = \Ts(s_1,t) = T_m$,  the Stefan condition at the melt front becomes 	
\begin{subequations} 
\begin{alignat}{2}
\left[{\q}^{\rm L} - {\q}^{\rm S}  \right] \cdot \n &= \rhos \left(u^*_1 - \usn \right) \left[ \left(\cpl - \cps\right)\left(T_m - T_r\right) + L_m + \frac{(\uln)^2}{2} - \frac{(\usn)^2}{2} - \frac{\pl}{\rhol} +  \frac{\ps}{\rhos}  \right] + \ps \usn - \pl \uln  \nonumber \\ 
& \qquad - \sigma_1 \mathcal{C}_1 u^*_1, \label{eqn_simplified3_energy_jump_LSa} \\
& =  \rhos \left(u^*_1 - \usn \right) \left[ \left(\cpl - \cps\right)\left(T_m - T_r\right) + L_m  + \frac{(\uln)^2}{2} - \frac{(\usn)^2}{2} \right] - \pl  \frac{\rhos}{\rhol} \left(u^*_1 - \usn \right) +  \ps u^*_1 \nonumber \\ 
& \qquad - \pl \uln - \sigma_1 \mathcal{C}_1 u^*_1.  \label{eqn_simplified3_energy_jump_LSb}
\end{alignat}
\end{subequations}

\noindent The jump condition for the momentum equation at the melt front yields the pressure difference $\ps - \pl$ 
\begin{align}\label{eqn_momentum1_jump_LS}
\left( \rhol \uln - \rhos \usn \right) u^*_1 & =
		\rhol \left(\uln\right)^2 - \rhos \left(\usn \right)^2 + \pl - \ps + \sigma_1 \mathcal{C}_1, \nonumber \\
\hookrightarrow \quad 	\ps - \pl &= \rhol \uln \left(\uln - u^*_1 \right) - \rhos \usn \left(\usn - u^*_1 \right) + \sigma_1 \mathcal{C}_1. 
\end{align}

\noindent Substituting $\ps$ from Eq.~\eqref{eqn_momentum1_jump_LS}, and the mass jump condition of Eq.~\eqref{eqn_mass_jump_LSb} into the energy jump  Eq.~\eqref{eqn_simplified3_energy_jump_LSb} we obtain
\begin{align}
\left[{\q}^{\rm L} - {\q}^{\rm S}  \right] \cdot \n &= 
\rhos \left(u^*_1 - \usn \right) \left[ \left(\cpl - \cps\right)\left(T_m - T_r\right) + L_m  + \frac{(\uln)^2}{2} - \frac{(\usn)^2}{2} \right] - \pl  \frac{\rhos}{\rhol} \left(u^*_1 - \usn \right) \nonumber \\ 
& \qquad  + \left[\pl + \rhol \uln \left(\uln - u^*_1 \right) - \rhos \usn \left(\usn - u^*_1 \right) + \cancel{\sigma_1 \mathcal{C}_1}\right] u^*_1 - 
		\pl \uln - \cancel{\sigma_1 \mathcal{C}_1 u^*_1}, \nonumber \\
     &=   \rhos \left(u^*_1 - \usn \right) \left[ \left(\cpl - \cps\right)\left(T_m - T_r\right) + L_m  + \frac{(\uln)^2}{2} - \frac{(\usn)^2}{2} \right]  - \pl  \frac{\rhos}{\rhol} \left(u^*_1 - \usn \right) \nonumber \\
     & \qquad + \pl \left(u^*_1 - \uln \right) +\rhos\left( \usn-u^*_1 \right) \left(\uln - \usn  \right) u^*_1. \label{eqn_simplified4_energy_jump_LS}
\end{align}

At this stage further simplifications to the right hand side of Eq.~\eqref{eqn_simplified4_energy_jump_LS} can be made by substituting $\usn$ and $\uln$ in terms of interface velocities $u_1^*$ and $u_2^*$ as derived in Eqs.~\eqref{eqn_uln} and~\eqref{eqn_usn}. This yields
\begin{subequations}
\begin{alignat}{2}
\left[{\q}^{\rm L} - {\q}^{\rm S}  \right] \cdot \n  &= \rhos \left(u^*_1 - \usn \right) \left[ \left(\cpl - \cps\right)\left(T_m - T_r\right) + L_m  + \frac{(\uln)^2}{2} - \frac{(\usn)^2}{2} + \left(\usn - \uln  \right) u^*_1\right]   \nonumber \\
& \qquad \underbrace{- \pl  \frac{\rhos}{\rhol} \left(\bluecancel{\frac{\rhol}{\rhos}u^*_1} - \left(\blackcancel{\frac{\rhol}{\rhos}}  - \redcancel{\frac{\rhov}{\rhos}} \right) u^*_2 \right) + \pl \left(\bluecancel{u^*_1} -\left( \blackcancel{1} - \redcancel{\frac{\rhov}{\rhol}} \right) u^*_2  \right)}_{= \, 0}, \\
&= \left(\rhol u^*_1 - \left(\rhol-\rhov \right)u^*_2 \right)\left[ \left(\cpl - \cps\right)\left(T_m - T_r\right) + L_m  + \underbrace{\frac{1}{2} \Biggl\{ \left( 1 - \frac{\rhov}{\rhol} \right) u^*_2 \Biggr\}^2}_{\textrm{kinetic energy}} \right. 
 \nonumber \\
& - \left. \underbrace{\frac{1}{2} \Biggl\{ \left( \frac{\rhol}{\rhos} - \frac{\rhov}{\rhos} \right) u^*_2 + \left( 1 - \frac{\rhol}{\rhos} \right)u^*_1 \Biggr\}^2+ \Biggl\{ \left( \frac{\rhol}{\rhos} - \frac{\rhov}{\rhos} \right) u^*_2 + \left( 1 - \frac{\rhol}{\rhos} \right)u^*_1- \left( 1 - \frac{\rhov}{\rhol} \right) u^*_2\Biggr\} u^*_1}_{\textrm{kinetic energy}}\right].   \label{eqn_stefan_ls_kinetic}
\end{alignat}
\end{subequations}

\noindent \textbf{\underline{Stefan conditions for the three phase Stefan problem:}} Omitting the kinetic energy (k.e.) terms in Eqs.~\eqref{eqn_simplified4_energy_jump_VL} and~\eqref{eqn_stefan_ls_kinetic}, which are dominant only during the initial periods of phase change, the Stefan condition at the vapor-liquid and liquid-solid interface reads as
\begin{subequations} \label{eqn_Stefan_conditions}
\begin{alignat}{2}
&\rhov \left[ \left( \cpv - \cpl \right) \left( T_v - T_r \right) + L_e \right] u^*_2 + \text{k.e. terms}
		= \left[ \kl \D{\Tl}{x} -\kv \D{\Tv}{x}   \right] \underset{x = s_2(t)}{\text{ }}, \label{eqn_Stefan_VL}\\
&\left(\rhol u^*_1 - \left(\rhol-\rhov \right)u^*_2 \right)\left[ \left(\cpl - \cps\right)\left(T_m - T_r\right) + L_m \right] + \text{k.e. terms} = \left[ \ks \D{\Ts}{x}  -\kl \D{\Tl}{x} \right]\underset{x = s_1(t)}{}. \label{eqn_Stefan_LS}
\end{alignat}
\end{subequations}
Note that when the density of all three phases is the same, then the kinetic energy terms in Eqs.~\eqref{eqn_simplified4_energy_jump_VL} and~\eqref{eqn_stefan_ls_kinetic} vanish, and the Stefan conditions of  Eqs.~\eqref{eqn_Stefan_conditions} become the standard ones as written in classical heat transfer textbooks (see for example~\cite{hahn2012heat}). 

\subsection{Similarity transformations} \label{sec_similarity_trans}
To find the evolution of temperature as a result of phase change, we need to solve the temperature equation for each phase. This is obtained by expressing the internal energy $e$ in terms of temperature, and neglecting terms related to pressure work and viscous dissipation in Eq.~\eqref{eqn_energy_cons}. The resulting temperature equations in the solid, liquid, and vapor phases are 
\begin{subequations}
\begin{alignat}{2}    
&\rhos \cps \left( \D{\Ts}{t} + \us \D{\Ts}{x} \right) &&= \ks \DD{\Ts}{x} \quad \in \Omegas(t), \label{eqn_T_sol} \\
&\rhol \cpl \left( \D{\Tl}{t} + \ul \D{\Tl}{x} \right) &&= \kl \DD{\Tl}{x} \quad \in \Omegal(t), \label{eqn_T_liq} \\
&\rhov \cpv \left( \D{\Tv}{t} + \uv \D{\Tv}{x} \right) &&= \kv \DD{\Tv}{x} \quad \in \Omegav(t). \label{eqn_T_vap} 
\end{alignat}
\end{subequations}
In Eqs.~\eqref{eqn_T_sol}-\eqref{eqn_T_vap}, the phase velocities can be expressed in terms of the interface velocities $u^*_1$ and $u^*_2$: $\uv \equiv 0$, $\ul \equiv \left(1 - \rrhovl\right)u^*_2$, and $\us \equiv \left(\rrhols - \rrhovs \right)u^*_2 + \left(1- \rrhols \right)u^*_1$.

Next, we demonstrate that all three temperature Eqs.~\eqref{eqn_T_sol}-\eqref{eqn_T_vap} admit a similarity transformation that converts these partial difference equations into ordinary ones. Specifically, instead of two variables, $x$ and $t$, the temperature in the three domains can be expressed in terms of a single variable $\eta^\Gamma$ as follows,
\begin{subequations}
\begin{alignat}{2}
\Ts(x,t) &= \Ts(\etas)   \qquad \text{with} \qquad \etas = \frac{x}{2\sqrt{\alphas t}} + c(t), \label{eqn_etas} \\
\Tl(x,t) &= \Tl(\etal)   \qquad \text{with} \qquad \etal = \frac{x}{2\sqrt{\alphal t}} + b(t), \label{eqn_etal} \\
\Tv(x,t) &= \Tv(\etav)   \qquad \text{with} \qquad \etav = \frac{x}{2\sqrt{\alphav t}}. \label{eqn_etav}
\end{alignat}
\end{subequations}
Here, $b(t)$ and $c(t)$ are functions of time that are yet to be determined. Interface positions $s_2(t)$ and $s_1(t)$ are also expressed in terms of two unknown functions of time, $\lambda(t)$ and $\beta(t)$, respectively, as follows,
\begin{subequations} \label{eqn_s1_s2}
\begin{alignat}{2}
s_1(t) = 2 \beta(t)  \sqrt{\alphas t}, \label{eqn_s1} \\		
s_2(t) = 2 \lambda(t)  \sqrt{\alphal t}.	\label{eqn_s2}	
\end{alignat}
\end{subequations}

\noindent Eqs.~\eqref{eqn_s1_s2} allow us to express the interface velocities in terms of $\lambda(t)$ and $\beta(t)$ as 
\begin{subequations} \label{eqn_u1_u2}
\begin{alignat}{2}
& u^*_1 = \dd{s_1}{t}= \beta \sqrt{\frac{\alphas}{t}}+2 \sqrt{\alphas t} \frac{\d \beta}{\d t},	\label{eqn_u1} \\
& u^*_2 = \dd{s_2}{t}= \lambda \sqrt{\frac{\alphal}{t}}+2 \sqrt{\alphal t} \frac{\d \lambda}{\d t}.
\end{alignat}
\end{subequations}

\noindent \textbf{\underline{Similarity transformation of Eq.~\eqref{eqn_T_sol} :}}

\noindent Inserting $\us$ from Eq.~\eqref{eqn_usn} into Eq.~\eqref{eqn_T_sol} yields
\begin{equation}
\rhos \cps \left( \D{\Ts}{t}  + \Bigl\{ \left(\rrhols - \rrhovs \right)u^*_2 + \left(1- \rrhols \right)u^*_1 \Bigr\} \D{\Ts}{x} \right) = \ks \DD{\Ts}{x}.  \label{eqn_T_sol_expanded}
\end{equation}
Rewriting  the derivatives of $\Ts$ in terms of $\etas$ we get
\begin{subequations} \label{eqns_sim_Ts_derivs}
\begin{alignat}{2}
\D{\Ts}{t} & =\dd{\Ts}{\etas}\D{\etas}{t} =\dd{\Ts}{\etas}\left(\frac{-x\left(\alphas t \right)^{-\frac{3}{2}}\alphas}{4}+\dd{c}{t}\right) = \dd{\Ts}{\etas}\left(-\frac{x}{4 t \sqrt{\alphas t}} + \dd{c}{t} \right), \label{eqn_sim_dTs/dt} \\
\D{\Ts}{x} &=  \dd{\Ts}{\etas} \frac{1}{2\sqrt{\alphas t}}, \label{eqn_sim_dTs/dx} \\
\DD{\Ts}{x} &= \frac{1}{4\alphas t} \frac{\d^2 \Ts}{\d{{\etas}^2}}.
\end{alignat}
\end{subequations}
Substituting Eqs.~\eqref{eqns_sim_Ts_derivs} into the solid phase temperature Eq.\eqref{eqn_T_sol_expanded}, we get
\begin{equation} \label{eqn_T_sol_expanded2}
\begin{split}
\rhos \cps \Bigg[ & \dd{\Ts}{\etas}\left(-\frac{x}{4 t \sqrt{\alphas t}} + \dd{c}{t} \right) + \Biggl\{\left(\rrhols- \rrhovs\right)\bigg(\lambda \sqrt{\frac{\alphal}{t}}+2 \sqrt{\alphal t} \dd{\lambda}{t}\bigg) \\ 
& +\left(1-\rrhols \right)\bigg(\beta \sqrt{\frac{\alphas}{t}}+2 \sqrt{\alphas t} \dd{\beta}{t}\bigg) \biggr\} \dd{\Ts}{\etas}\frac{1}{2\sqrt{\alphas t}} \Bigg] =  \frac{\ks}{4\alphas t} \frac{\d^2 \Ts}{\d{{\etas}^2}}. 
\end{split}
\end{equation}
By using the definition of $\etas$ from Eq.~\eqref{eqn_etas}, we further simplify Eq.~\eqref{eqn_T_sol_expanded2} to 
\begin{equation}\label{eqn_sim_Ts_expanded3}
\frac{\d^2 \Ts}{\d{{\etas}^2}}-\dd{\Ts}{\etas}\Bigg[-2\etas + \underbrace{2c +4t \dd{c}{t}+\left(\rrhols - \rrhovs\right)\biggl\{2\lambda \sqrt{\frac{\alphal}{\alphas}}+4t\sqrt{\frac{\alphal}{\alphas}} \dd{\lambda}{t}\biggr\}+\left(1-\rrhols \right)\biggl\{2\beta+4t \frac{\mathrm{d} \beta}{\mathrm{d} t}\biggr\}}_{\textrm{extraneous coefficient}}\Bigg]=0.
\end{equation}

\noindent For the similarity transformation of Eq.~\eqref{eqn_sim_Ts_expanded3} to exist, we require the highlighted extraneous coefficient of  $ \dd{\Ts}{\etas} $ to vanish. This can be achieved by choosing the function $c(t)$ to be of the form
\begin{equation}\label{eqn_c(t)}
c(t)=-\lambda(t) \sqrt{\frac{\alphal}{\alphas}} \left(\rrhols - \rrhovs \right)-\beta(t) \left(1-\rrhols \right).
\end{equation}
With the above choice of $c(t)$, the similarity transformation of the solid phase temperature Eq.~\eqref{eqn_sim_Ts_expanded3} reads as
\begin{align}
\frac{\d^2 \Ts}{\d{{\etas}^2}} +2\etas \dd{\Ts}{\etas}=0  \quad \text{with} \quad \etas = \frac{x}{2\sqrt{\alphas t}} -\lambda(t)\sqrt{\frac{\alphal}{\alphas}} \left(\rrhols - \rrhovs \right)-\beta(t)\left(1 - \rrhols\right). \label{eqn_sim_Ts}
\end{align}


\noindent \textbf{\underline{Similarity transformation of Eq.~\eqref{eqn_T_liq} :}} 

\noindent Inserting $\ul$ from Eq.~\eqref{eqn_uln} into the liquid phase temperature Eq.~\eqref{eqn_T_liq} gives
\begin{equation}
\rhol \cpl \left( \D{\Tl}{t}  + \Bigl\{ \left(1 - \rrhovl\right) u^*_2 \Bigr\} \D{\Tl}{x} \right) = \kl \DD{\Tl}{x}.  \label{eqn_T_liq_expanded}
\end{equation}
Rewriting the derivatives of $\Tl$ in terms of $\etal$, we get
\begin{subequations} \label{eqns_sim_Tl_derivs}
\begin{alignat}{2}
\D{\Tl}{t} & =\dd{\Tl}{\etal}\D{\etal}{t} =\dd{\Tl}{\etal}\left(\frac{-x\left(\alphal t \right)^{-\frac{3}{2}}\alphal}{4}+\dd{b}{t}\right) = \dd{\Tl}{\etal}\left(-\frac{x}{4 t \sqrt{\alphal t}} + \dd{b}{t} \right), \label{eqn_sim_dTl/dt} \\
\D{\Tl}{x} &=  \dd{\Tl}{\etal} \frac{1}{2\sqrt{\alphal t}}, \label{eqn_sim_dTl/dx} \\
\DD{\Tl}{x} &= \frac{1}{4\alphal t} \frac{\d^2 \Tl}{\d{{\etal}^2}}.
\end{alignat}
\end{subequations}
Substituting Eqs.~\eqref{eqns_sim_Tl_derivs} into the liquid phase temperature Eq.\eqref{eqn_T_liq_expanded}, we get
\begin{equation} \label{eqn_T_liq_expanded2}
\begin{split}
\rhol \cpl \Bigg[ \dd{\Tl}{\etal}\left(-\frac{x}{4 t \sqrt{\alphal t}} + \dd{b}{t} \right) + \left(1-\rrhovl \right)\left(\lambda \sqrt{\frac{\alphal}{t}}+2 \sqrt{\alphal t} \dd{\lambda}{t}\right) \dd{\Tl}{\etal}\frac{1}{2\sqrt{\alphal t}} \Bigg] = \frac{\kl}{4\alphal t} \frac{\d^2 \Tl}{\d{{\etal}^2}},
\end{split}
\end{equation}
which can be further simplified by using the definition of $\etal$ from Eq.~\eqref{eqn_etal} to
\begin{equation} \label{eqn_T_liq_expanded3}
\frac{\d^2 \Tl}{\d{{\etal}^2}} - \dd{\Tl}{\etal} \left[-2\etal + \underbrace{2b + 4t \dd{b}{t} +\left(2\lambda + 4t\dd{\lambda}{t}\right)\left(1- \rrhovl\right)}_{\textrm{extraneous coefficient}}\right] =0.
\end{equation}
Eq.~\eqref{eqn_T_liq_expanded3} admits a similarity transformation if we choose $b(t)$ as
\begin{equation} \label{eqn_b(t)}
 b(t) =-\lambda(t)\left(1- \rrhovl \right).
\end{equation}
With this choice of $b(t)$, the extraneous coefficient of $\d \Tl/ \d \etal$ in Eq.~\eqref{eqn_T_liq_expanded3} vanishes.  The similarity transformation of the liquid phase temperature Eq.~\eqref{eqn_T_liq_expanded3} reads as
\begin{align}
\frac{\d^2 \Tl}{\d{{\etal}^2}} +2\etal \dd{\Tl}{\etal}=0  \quad \text{with} \quad \etal = \frac{x}{2\sqrt{\alphal t}} -\lambda(t)\left(1- \rrhovl \right). \label{eqn_sim_Tl}
\end{align}


\noindent \textbf{\underline{Similarity transformation of Eq.~\eqref{eqn_T_vap} :}} 

\noindent Setting $\uv = 0$ in the vapor phase temperature Eq.~\eqref{eqn_T_vap} yields
\begin{equation}
\rhov \cpv \D{\Tv}{t} = \kv \DD{\Tv}{x}.  \label{eqn_T_vap_expanded}
\end{equation}
Following similar (but less involved) steps, the similarity transformation of Eq.~\eqref{eqn_T_vap_expanded} reads as 
\begin{align}
\frac{\d^2 \Tv}{\d{{\etav}^2}} +2\etav \dd{\Tv}{\etav}=0  \quad \text{with} \quad \etav = \frac{x}{2\sqrt{\alphav t}}. \label{eqn_sim_Tv}
\end{align}


\subsection{Similarity solutions} \label{sec_similarity_trans}

Integrating the transformed equations~\eqref{eqn_sim_Ts},~\eqref{eqn_sim_Tl}, and~\eqref{eqn_sim_Tv} twice with respect to $\etas$, $\etal$, and $\etav$, respectively, we obtain the following general solutions for the temperature in the three domains:
\begin{subequations} \label{eqn_sim_solns}
\begin{alignat}{2}
 &\Ts(x,t) = \Ts(\etas)= B_1+A_1 \, \text{erfc}\left(\frac{x}{2\sqrt{\alphas t}}-\lambda(t)\sqrt{\frac{\alphal}{\alphas}}\left(\rrhols - \rrhovs \right)-\beta(t)\left(1 - \rrhols\right)\right), \label{eqn_sim_Ts_soln} \\
 &\Tl(x,t) = \Tl(\etal) = B_2 + A_2 \, \text{erf}\left( \frac{x}{2\sqrt{\alphal t}}-\lambda(t)\left(1 - \rrhovl\right)\right), \label{eqn_sim_Tl_soln} \\
 &\Tv(x,t) = \Tv(\etav) = B_3+ A_3 \, \text{erf}\left( \frac{x}{2\sqrt{\alphav t}}\right). \label{eqn_sim_Tv_soln}
\end{alignat}    
\end{subequations}
Six boundary conditions are needed to determine the six unknown coefficients: $A_1$, $A_2$, $A_3$, $B_1$, $B_2$, and $B_3$. These include four temperature conditions at $x = 0$, $x = l$, $x = s_1(t)$, and $x = s_2(t)$, and two Stefan conditions at $x = s_1(t)$, and $x = s_2(t)$.  

Applying the temperature boundary condition $T(x = 0,t) = \Tv(\etav = 0) = T_\infty$, and $T(x = l\rightarrow \infty,t) = \Ts(\etas\rightarrow \infty) = T_o$ yields
\begin{equation}
B_3 = T_\infty \quad \text{and} \quad B_1 = T_o.
\end{equation}

Using the saturation temperature conditions for the two interfaces $ T(s_2,t)=T_v $ and $ T(s_1,t)=T_m $, we determine the coefficients $ A_3 $ and $ A_1 $ based on the similarity solutions given by Eqs.~\eqref{eqn_sim_Tv_soln} and~\eqref{eqn_sim_Ts_soln}, respectively, as
\begin{equation}\label{eqn_A1_A3}
A_3=\frac{T_v-T_\infty}{\text{erf}\left(\lambda(t)\sqrt{\frac{\alphal}{\alphav}}\right)} \quad \text{and} \quad A_1=\frac{T_m-T_o}{\text{erfc}\left(\beta(t)\rrhols -\lambda(t)\sqrt{\frac{\alphal}{\alphas}}\left( \rrhols - \rrhovs \right) \right)}.
\end{equation}
In the above, we made use of Eqs.~\eqref{eqn_s1_s2} to simplify the coefficients. Note that the coefficients $A_1\left(\lambda(t)\right)$ and $A_3\left(\lambda(t),\beta(t)\right)$ are implicit functions of time. If $\lambda$ and $\beta$ are time-(in)dependent, $A_1$ and $A_3$ are as well.  

Using the saturation temperature condition at the two interfaces bounding the liquid phase, $ T(s_2,t)=T_v $ and $ T(s_1,t)=T_m $, we can determine the coefficients $ B_2$ and $ A_2$ based on the similarity solution given by Eq.~\eqref{eqn_sim_Tl_soln} as follows:
\begin{equation}\label{eqn_A2_B2}
A_2=\frac{T_m-T_v}{\text{erf}\left(\beta(t)\sqrt{\frac{\alphas}{\alphal}}-\lambda(t)\left(1 - \rrhovl\right)\right)-\text{erf}\left( \lambda(t) \rrhovl\right)} \quad \text{and} \quad B_2 = T_v - A_2\, \text{erf}\left(\lambda(t) \rrhovl\right).
\end{equation}

The remaining unknowns, $\lambda(t)$ and $\beta(t)$, are determined by substituting the similarity solutions (Eqs.~\eqref{eqn_sim_solns}) into the two Stefan conditions  (Eqs.~\eqref{eqn_Stefan_conditions}),
\begin{subequations}\label{eqn_coupled_stefan}
\begin{alignat}{2}
 &\rhov \left[ \left( \cpv - \cpl \right)(T_v - T_r) + L_e \right] \lambda\sqrt{\alphal} = \frac{\kl}{\sqrt{\pi\alphal}}\frac{T_m-T_v}{\text{erf}\left(\beta\sqrt{\frac{\alphas}{\alphal}}-\lambda\left(1 - \rrhovl\right)\right)-\text{erf}\left( \lambda \rrhovl\right)}   \exp\left(-\left(\lambda \rrhovl \right)^2\right) \nonumber \\
		& \qquad \qquad -\frac{\kv}{\sqrt{\pi\alphav}}\frac{T_v-T_\infty}{\text{erf}\left(\lambda\sqrt{\frac{\alphal}{\alphav}}\right)}\exp\left(-\lambda^2\frac{\alphal}{\alphav}\right) + \underbrace{\mathcal{O}\bigg( \frac{\lambda^3}{t} + \left(\dd{\lambda}{t}\right)^3 + \cdots \bigg)}_{\text{kinetic energy}},       \\
&\left(\rhol\beta \sqrt{\alphas} -(\rhol - \rhov)\lambda \sqrt{\alphal} \right)\left\{ (\cpl - \cps)(T_m - T_r) + L_m \right\} =  \nonumber \\
& \qquad \qquad -\frac{\kl}{\sqrt{\pi \alphal}} \frac{T_m-T_v}{\text{erf}\left(\beta\sqrt{\frac{\alphas}{\alphal}}-\lambda\left(1 - \rrhovl\right)\right)-\text{erf}\left( \lambda \rrhovl\right)}   \exp\left( -\left[\beta\sqrt{\frac{\alphas}{\alphal}}-\lambda\left(1 - \rrhovl\right) \right]^2\right)  \nonumber \\
& \qquad \qquad -\frac{\ks}{\sqrt{\pi \alphas}} \frac{T_m-T_o}{\text{erfc}\left(\beta\rrhols -\lambda\sqrt{\frac{\alphal}{\alphas}}\left( \rrhols - \rrhovs \right) \right)} \exp\left(-\left[ \beta \rrhols -\lambda \sqrt{\frac{\alphal}{\alphas}}(\rrhols - \rrhovs)\right]^2 \right) \nonumber \\
 &\qquad \qquad + \underbrace{\mathcal{O}\bigg( \frac{\lambda^2 \beta}{t} + \left(\dd{\lambda}{t}\right)^2\left(\dd{\beta}{t}\right) + \cdots \bigg)}_{\text{kinetic energy}}.
\end{alignat}
\end{subequations}

If we omit the kinetic energy terms from Eqs.~\eqref{eqn_coupled_stefan}, then we have coupled, time-independent transcendental equations in $\lambda$ and $\beta$. Accordingly, the solution to the coupled Eqs.~\eqref{eqn_coupled_stefan} yields time-independent values of $\lambda$ and $\beta$, which also result in time-independent values of $A_1$, $A_2$, $A_3$, and $B_2$. In this work, we neglect the kinetic energy terms while solving the coupled transcendental Eqs.~\eqref{eqn_coupled_stefan}.

\section{A sharp-interface technique to solve the three-phase Stefan problem numerically} \label{sec_sharp_interface_method}

In this section, we present a fixed-grid, sharp-interface method to solve the heat/temperature Eqs.~\eqref{eqn_T_sol}-\eqref{eqn_T_vap} in the evolving solid, liquid and vapor domains. Stefan conditions and saturation temperature conditions on evolving vapor-liquid and liquid-solid interfaces are imposed sharply in this approach. 

The one-dimensional domain $\Omega: = 0 \le x \le l$ is discretized into $N$ cells of uniform size $\Delta x = l/N$. The discrete temperature $T_i$ and velocity $u_i$ are both stored at the $i^{\rm th}$ (with $i = 1,\ldots, N$) cell-center having a coordinate $x_i = (i - \half) \Delta x$. 

The heat equation is discretized using a second-order finite difference stencil. We treat conduction/diffusion implicitly and convection explicitly. The heat equation is discretized only for regular grid cells that do not contain or abut the moving interface. In contrast, irregular cells containing or abutting the interface impose a saturation temperature condition. With $n$ denoting the time level, and $\Delta t$ denoting the time-step size, the discrete form of the heat equation for a regular grid cell $i$ at (midpoint) time $t = (n+\half)\Delta t$ reads as
\begin{equation} \label{eqn_discrete_heat_reg}
\frac{T_i^{n+1} - T_i^n}{\Delta t} + \bigg(u_i \frac{\widetilde{T}_{i+\half} - \widetilde{T}_{i-\half}}{\Delta x} \bigg)^{n+\half}  = \kappa \bigg[ \frac{T_{i+1}^{n+1} - 2T_{i}^{n+1} + T_{i-1}^{n+1}}{2(\Delta x)^2} +  \frac{T_{i+1}^{n} - 2T_{i}^{n} + T_{i-1}^{n}}{2(\Delta x)^2}\bigg].
\end{equation}
Here, $\widetilde{T}$ denotes the CUI (cubic upwind interpolated)~\cite{nangia2019robust} limited temperature field on the left ($i - \half$) and right ($i+\half$) faces of the regular grid cell $i$. An extrapolation of the form $\phi^{n+\half} = \threehalf \phi^n - \half \phi^{n-1}$ is performed to evaluate the velocity and temperature at time level $n+\half$ for the convective term. 

\begin{figure}[]
	\begin{center}
		\includegraphics[scale=0.5]{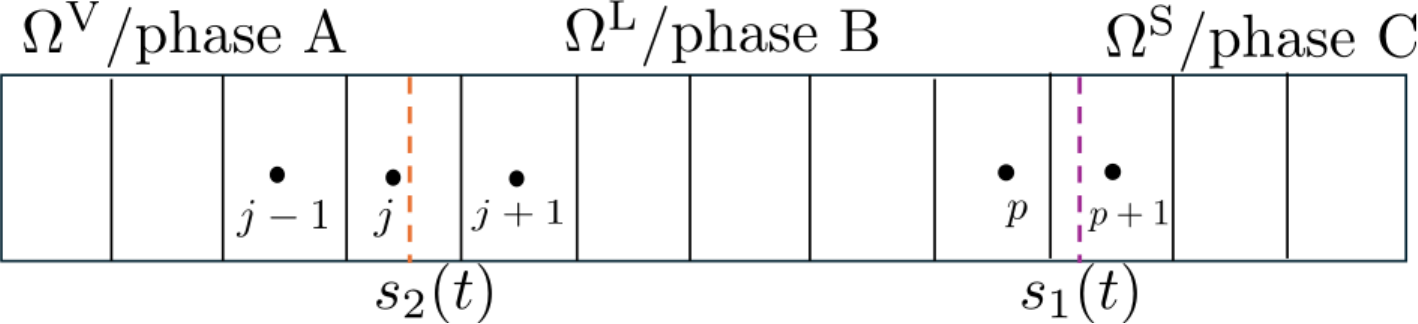}
	\end{center}
	\caption{Schematic of the 1D grid used to discretize the heat equation using the sharp interface technique. Irregular cells $j$ and $j+1$, $p$ and $p+1$ abut the interfaces separating phases $A$ and $B$, and phases $B$ and $C$, respectively. The cell centers are marked with $(\bullet)$, and the two evolving interfaces are marked with dashed lines (\texttt{---}). }\label{fig_grid_sharp_interface}
\end{figure}

Grid cells adjacent to the interface, say $j$ and $j+1$, are indicated as irregular in Fig.~\ref{fig_grid_sharp_interface}. As mentioned before, we do not discretize the heat equation for irregular cells. Thus, to complete the system of equations for $N$ unknown temperatures, we need two additional equations. By associating cells $j-2$, $j-1$, and $j$ with phase $A$ and cells $j+1$, $j+2$, and $j+3$ with phase $B$, we can calculate the temperature at the interface from each side by using one-sided quadratic extrapolation. One-sided extrapolated temperature at the interface is equated to saturation temperature $T_{\rm sat}$ (i.e., $T_m$ or $T_v$) to close the system of equations:
\begin{subequations} \label{eqn_extrap_temp}
 \begin{alignat}{2}
 &T_A \bigg|_{x = s} =  a_1 T_{A,j} + a_2 T_{A,j-1} + a_3 T_{A,j-2} = T_{\rm sat}, \\
 &T_B \bigg|_{x = s} =  b_1 T_{B,j+1} + b_2 T_{B,j+2} + b_3 T_{j,j+3} = T_{\rm sat}.
 \end{alignat}
\end{subequations}
The coefficients $a_k$ and $b_k$ appearing in Eqs.~\eqref{eqn_extrap_temp} are determined by fitting a one-dimensional, second-order Lagrange polynomial to the neighboring nodes. Taking phase $A$ as an example, the coefficients $a_k$ read as follows
\begin{subequations}\label{eqn_coef_ak}
\begin{alignat}{2}
	&	a_1=\frac{\left(s - x_{j-1}\right)\left(s - x_{j-2}\right)}{\left(x_j-x_{j-1} \right)\left(x_{j} -x_{j-2}\right)}, \\
		&  a_2=\frac{\left(s-x_j\right)\left(s-x_{j-2}\right)}{\left(x_{j-1}-x_j\right)\left(x_{j-1}-x_{j-2}\right)}, \\
		&  a_3=\frac{\left(s-x_j\right)\left(s-x_{j-1}\right)}{\left(x_{j-2}-x_j\right)\left(x_{j-2}-x_{j-1}\right)}.
\end{alignat}
\end{subequations}
A similar process is followed for irregular cells $p$ and $p+1$ that abut the second interface.  

The discrete Eqs.~\eqref{eqn_discrete_heat_reg} and~\eqref{eqn_extrap_temp} form a system of $N$ equations that are inverted using a direct solver to calculate the $N$ unknown temperatures. At this stage, only the velocities of the two interfaces, $u_1^{*,n+1}$ and $u_2^{*,n+1}$ need to be determined. Those can be calculated by evaluating the one-sided derivatives of temperature $T^{n+1}$ required for the right hand side of Stefan Eqs.~\eqref{eqn_Stefan_conditions}. It is possible to express the one-sided derivative of temperature at the interface in terms of neighboring nodal temperatures in phase $A$ as
\begin{subequations} \label{eqn_taylor_series_dtdx}
\begin{alignat}{2} 
    \dd{T_A}{x}\Bigg|_{x = s} &=  \sum_{j = 1}^{3}c_{A,j} T_{A,j}, \label{eqn_taylor_series_dtdxa}\\
    \hookrightarrow \dd{T_A}{x}\Bigg|_{x = s} &= \sum_{j = 1}^{3} c_{A,j}\left(T_A\bigg|_s + \dd{T_A}{x}\bigg|_s (x_j - s) + \half \dD{T_A}{x}\bigg|_s (x_j - s)^2 + \mathcal{O}\left((x_j - s)^3\right) \right). \label{eqn_taylor_series_dtdxb} 
\end{alignat}
\end{subequations}
Eq.~\eqref{eqn_taylor_series_dtdxb} follows from Eq.~\eqref{eqn_taylor_series_dtdxa} through the use of the Taylor series expansion to express neighboring nodal temperatures in terms of the interface temperature and its derivatives. Ignoring higher order derivatives in Eq.~\eqref{eqn_taylor_series_dtdx}, and equating coefficients of $T_A$, $\dd{T_A}{x}$, and $\dD{T_A}{x}$ on both sides of the equation, yields a $3\times3$ system of equations for the coefficients $c_{A,j}$
 \begin{equation}
 	\begin{bmatrix}
 		1 & 1 & 1 \\
 		\Delta x_1 & \Delta x_2 & \Delta x_3 \\
 		(\Delta x_1)^2 & (\Delta x_2)^2 & (\Delta x_3)^2 \\
 	\end{bmatrix}
 	\begin{bmatrix}
 		c_{A,1} \\
 		c_{A,2} \\
 		c_{A,3} \\
 	\end{bmatrix}
 	=
 	\begin{bmatrix}
 		0 \\
 		1  \\
 		0 \\
 	\end{bmatrix}.
 \end{equation}
 Here, $\Delta x_j = (x_j - s)$. To calculate $\dd{T_B}{x}\bigg|_{x = s}$ for phase $B$, a similar procedure is followed. 

\section{Software implementation}
All MATLAB- and Mathematica-based codes used to generate results in this work can be found on the GitHub repository~\url{https://github.com/amneetb/ThreePhaseStefan}. 

\section{Results}

In this section, we solve both two- and three-phase Stefan problems. While the current work focuses on the latter problem, the former problem is simpler and well studied in the literature. We use it as a reference for validating the numerical method proposed in Sec.~\ref{sec_sharp_interface_method}. The PCM's thermophysical properties are largely based on aluminum metal. For the solid, liquid, and vapor phases of the PCM, the thermophysical properties are listed in Table~\ref{tab_thermophys_properties}. These values are taken from references~\cite{thirumalaisamy2023low,doble2007perry,hatch1984aluminium,desai1987thermodynamic}. 

\begin{table}[]
\centering
\caption{Thermophysical properties of the PCM (largely aluminum-based) used to simulate the two- and three-phase Stefan problems.}
\label{tab_thermophys_properties}
\begin{tabular}{ll l}
Property & Value & Units\\
\midrule
Density of solid $\rhos$                  & 2698.72  & kg/m$^3$ \\
Density of liquid $\rhol$                 & 2368  & kg/m$^3$ \\
Density of vapor $\rhov$                  & 0.08644 & kg/m$^3$ \\
Thermal conductivity of solid   $\ks$     &   211  & W/m.K  \\
Thermal conductivity of liquid  $\kl$     & 91   & W/m.K    \\
Thermal conductivity of vapor  $\kv$      & 115.739  &W/m.K    \\
Specific heat of solid  $\cps$            & 910 & J/kg.K      \\
Specific heat of liquid  $\cpl$           & 1042.4 & J/kg.K   \\
Specific heat of vapor  $\cpv$            & 770.69 &  J/kg.K   \\
Melting temperature  $T_m$                & 933.6  & K     \\
Vaporization temperature  $T_v$           &  2767  & K     \\
Latent heat of melting $L_m$              &  383840 & J/kg    \\
Latent heat of vaporization $L_e$         &  9462849.518 & J/kg    \\
\bottomrule
 \end{tabular}
\end{table}

\subsection{Two-phase Stefan problem: melting of a solid PCM}\label{sec_2pS_ana_num}

\begin{figure}[]
	\begin{center}
		\includegraphics[scale=0.4]{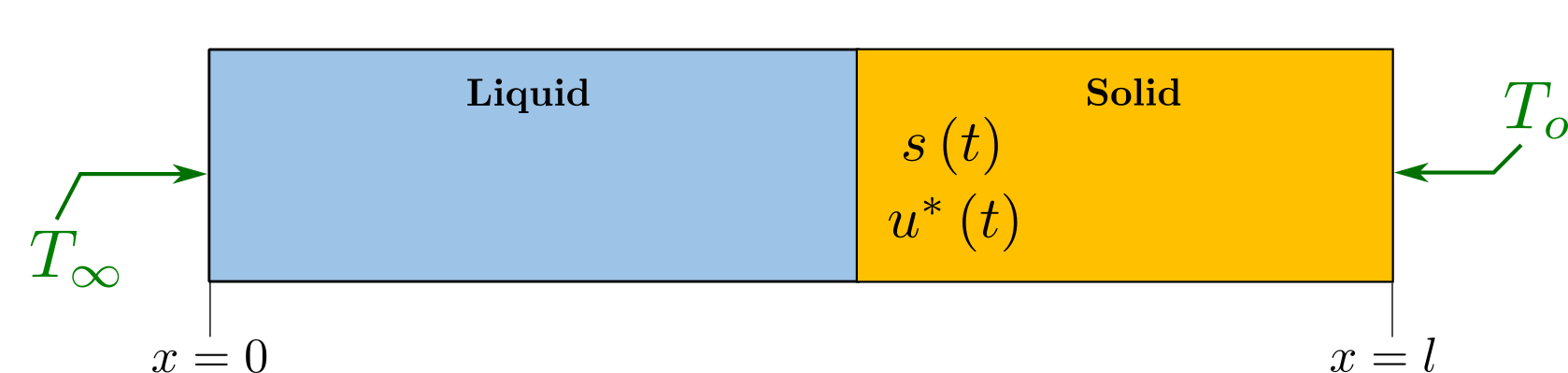}
	\end{center}
	\caption{Schematic of the two-phase Stefan problem illustrating melting of an initially solid PCM due to an imposed temperature condition at $x = 0$.}\label{fig_2pStefan}
\end{figure}

The PCM occupies a one-dimensional domain $\Omega: 0 \leq x \leq l$, with the solid phase originally occupying the entire domain at $t = 0$. The initial temperature is uniform and below the melting temperature $(T_i < T_m)$. The initial temperature is taken to be $T_i = 298$ K. The right boundary is open, while the left boundary is closed. As the length ($l = 1$ m) of the domain is large, the temperature at the right boundary remains at the initial temperature of the PCM, i.e., $T_o = T_i$. Imposing a temperature of $T_\infty = 2200$ K at the left boundary causes the PCM to melt ($T_m = 933.6$ K), and the liquid-solid melt front moves towards the right boundary, as illustrated in Fig.~\ref{fig_2pStefan}. The remainder of the thermophysical properties are taken from Table~\ref{tab_thermophys_properties}.

The analytical solution of this problem reads as~\cite{thirumalaisamy2023low}:
\begin{subequations} \label{eqn_2p_stefan_ana}
\begin{alignat}{2}
&\textbf{Solid temperature:}  &&\Ts(x,t) = T_o + a\,\text{erfc}\left(\frac{x}{2\sqrt{\alphas t}} - \beta\left(1-\rrhols\right)\right), \label{eqn_Tprofile_solid_2p} \\
&\textbf{Liquid temperature:}  && \Tl(x,t) =  T_\infty + b\,\text{erf}\left(\frac{x}{2\sqrt{\alphal t}}\right), \label{eqn_Tprofile_liquid_2p} \\
& &&  a = \frac{T_m-T_o}{\text{erfc}\left( \beta \rrhols \right) } \quad \text{and}   \quad b =  \frac{T_m-T_\infty}{\text{erf}\left(\beta \sqrt{\frac{\alphas}{\alphal}} \right)}, \\
&\textbf{Interface position and velocity:}  \quad && s(t) = 2\beta \sqrt{\alphas t} \quad \text{and} \quad u^* = \beta \sqrt{\frac{\alphas}{t}}, \label{eqn_s_u_2p}\\
&\textbf{Transcendental equation:} \quad && \rhol \left[ \left( \cpl - \cps \right)(T_m - T_r) + L_m \right] \beta\sqrt{\alphas}= \nonumber \\
& && -\ks\frac{T_m-T_o}{\text{erfc}\left(\beta \rrhols  \right)}\frac{\exp\left(-\left(\beta \rrhols  \right)^2\right)}{\sqrt{\pi\alphas}}  -\kl \frac{T_m-T_\infty}{\text{erf}\left(\beta\sqrt{\frac{\alphas}{\alphal}}\right)}\frac{\exp\left(-\beta^2\frac{\alphas}{\alphal}\right)}{\sqrt{\pi\alphal}}.	
\end{alignat}
\end{subequations}
In Eqs.~\eqref{eqn_2p_stefan_ana}, we have ignored the (non-dominant) kinetic energy terms, which were considered in our prior work. 

\begin{figure}[]
\label{result-melting}
    \begin{center}
        \subfigure[]{\includegraphics[scale=0.18]{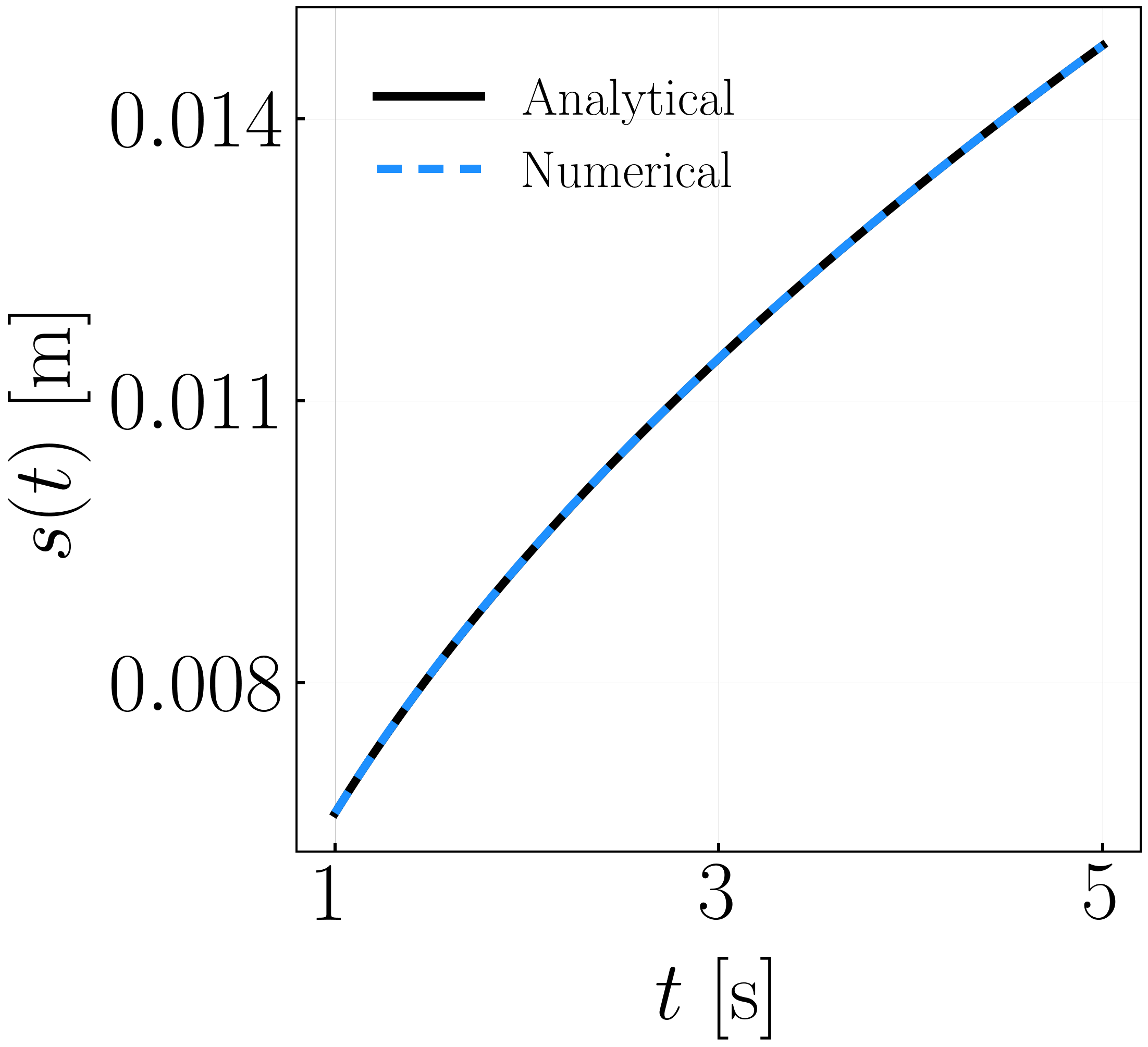}
        \label{fig_2p_A}
        }
        \hspace*{0.5cm}
        \subfigure[]{\includegraphics[scale=0.18]{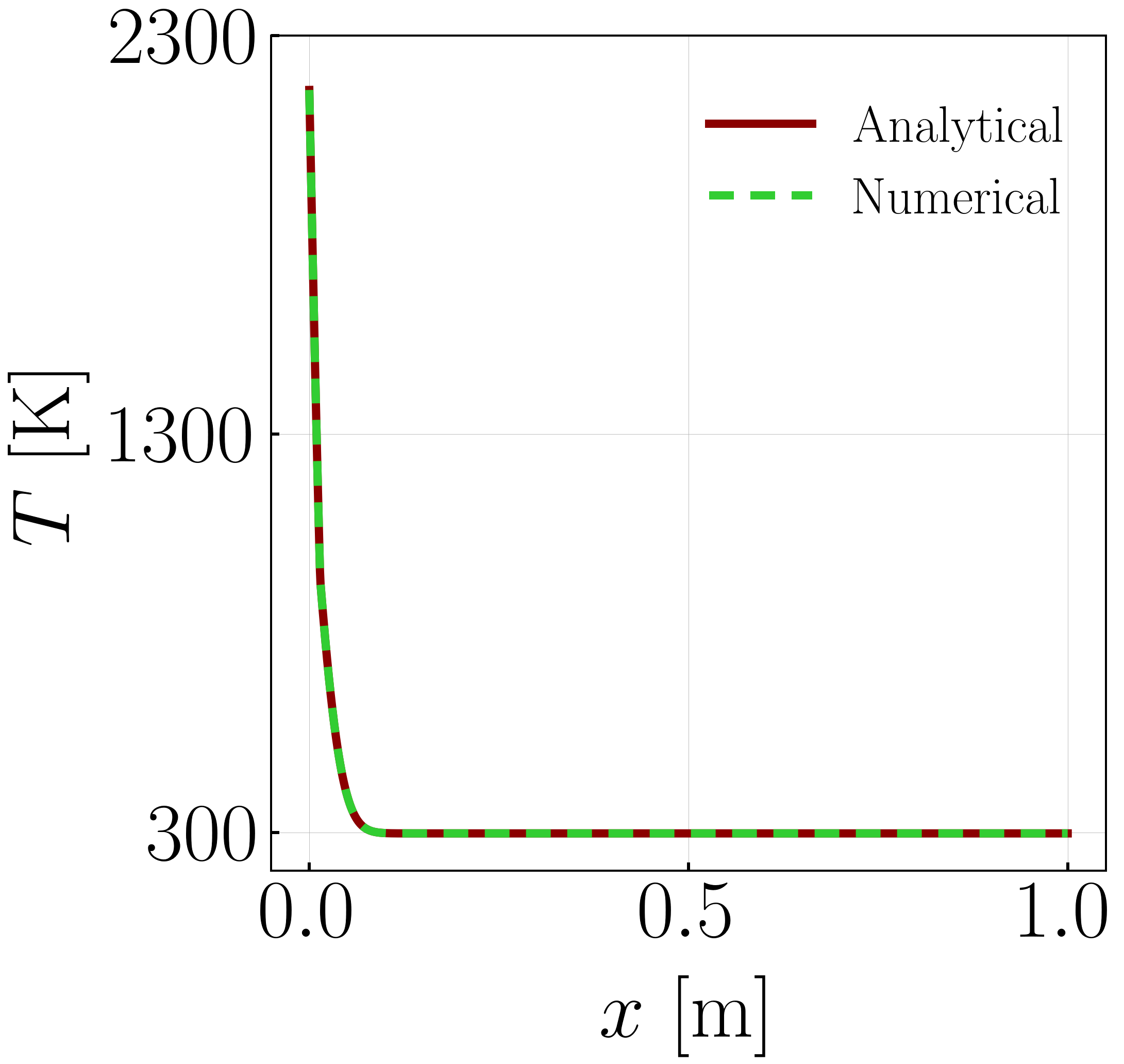}
        \label{fig_2p_B}
        }

        \subfigure[]{\includegraphics[scale=0.18]{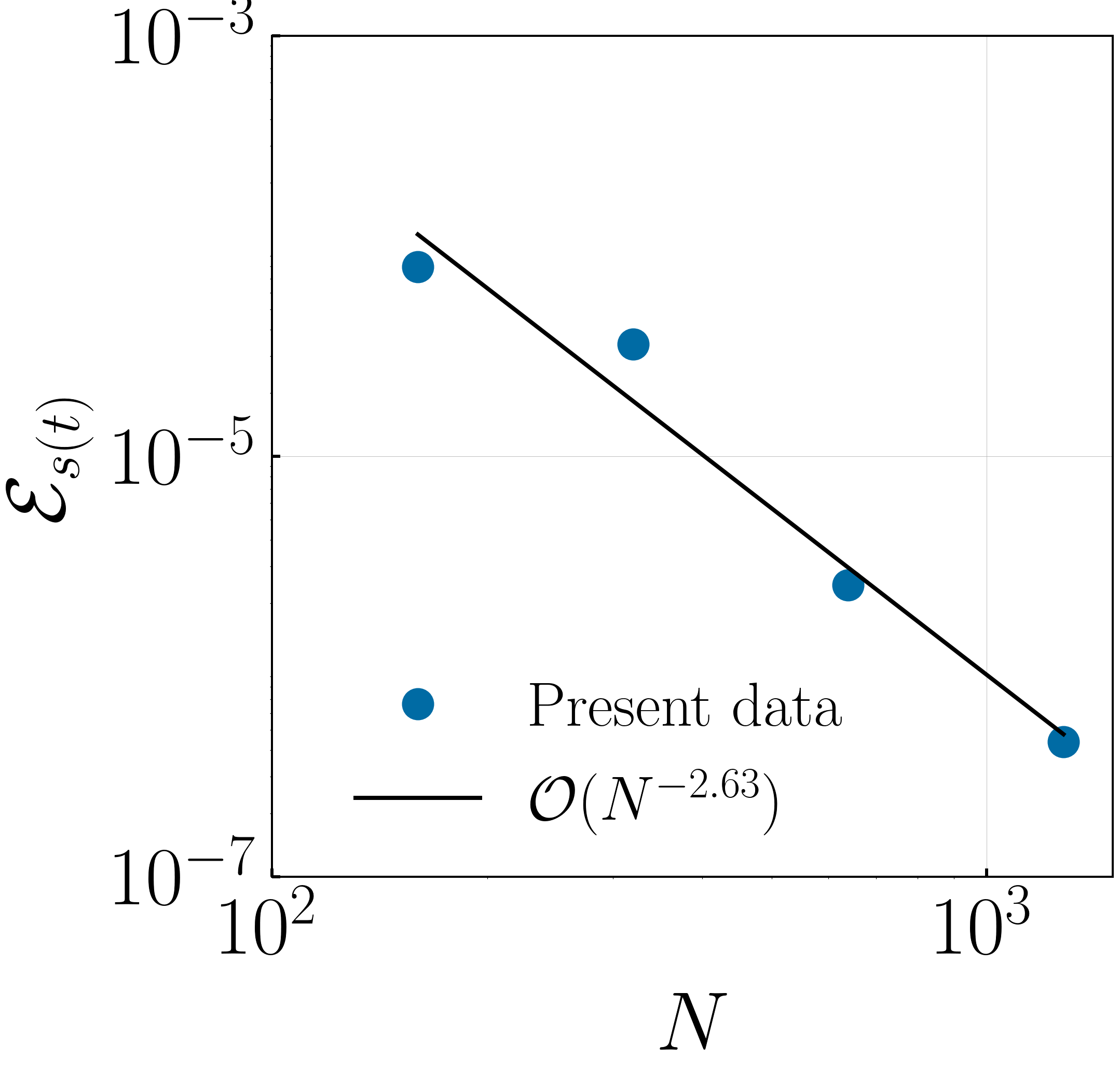}
        \label{fig_2p_C}
        }
         \hspace*{0.5cm}
        \subfigure[]{\includegraphics[scale=0.18]{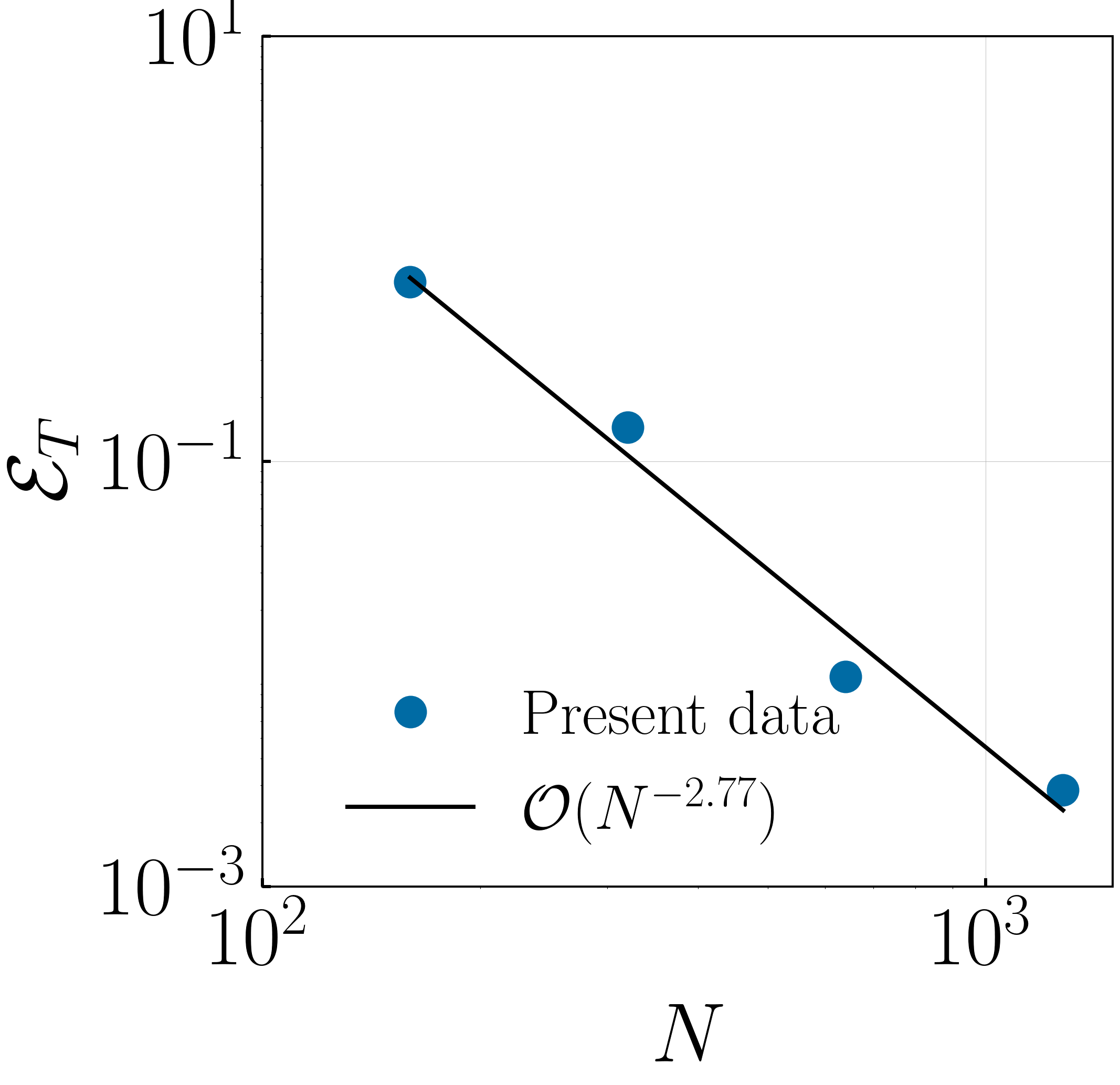}
        \label{fig_2p_D}
        }
        \end{center}
        \caption{Comparison of numerical results against the analytical solutions for the two-phase Stefan problem for \subref{fig_2p_A} interface position and \subref{fig_2p_B} temperature distribution across the domain. Convergence rate of the numerical error for \subref{fig_2p_C} interface position and \subref{fig_2p_D} temperature distribution as a function of grid size $N$.}
\end{figure}

To solve the two-phase Stefan problem numerically, we initialize the interface position and temperature in the domain based on the analytical solution at $t = 1$ s. By doing so, at least three grid cells are placed in the liquid phase to compute the gradient of temperature and to impose a saturation temperature condition on the liquid side of the interface. This is the main limitation of a sharp-interface approach, which requires both/all phases to be present within the domain. Our recently introduced low Mach enthalpy method~\cite{thirumalaisamy2023low,thirumalaisamy2025consistent}, which is a diffuse interface method, does not suffer from this limitation. Extending the low Mach method to solve three-phase Stefan problems with MSNBC remains a future task. The computational domain is discretized into $N = 1280$ cells. The time step size is taken to be $\Delta t = c \Delta x/u^*$, in which $c = 0.005$ represents the CFL number. The interface velocity $u^*$ is evaluated at $t = 1$ s, where it is at its maximum in the considered time period according to Eq.~\eqref{eqn_s_u_2p}. The numerical simulation is run from $t = 1$ to $t = 5$ s. Figs.~\ref{fig_2p_A} and~\ref{fig_2p_B} compare the analytical and numerical solutions for the interface position and temperature in the domain, respectively. An excellent agreement is found between the two. 

To determine the order of accuracy of the numerical method, we conducted simulations on four grids of size: $N = \{ 160,320,640,1280\}$. The CFL number is fixed at $c = 0.005$ for all grids. On each grid the simulation is run from $t = 6 $ to $t = 8$ s. The $L^2$-norm of error between the numerical and analytical solution is calculated for interface position and temperature distribution in the domain. The former is computed for the entire simulation run, whereas the latter is computed at the end of the simulation. The $L^2$ error for a quantity $\psi$ is defined as the root mean squared error (RMSE) of the vector $||\mathcal{E}_\psi||_{\rm RMSE} = ||\psi_{\rm analytical} - \psi_{\rm numerical}||_2 / \sqrt{\mathcal{S}}$. Here, $\mathcal{S}$ represents the size of the vector $\mathcal{E}_\psi$. Figs.~\ref{fig_2p_C} and~\ref{fig_2p_D} plot the actual error data as filled markers and best fit lines to determine the convergence rate of the error as a function of grid size $N$. It can be observed that the method converges with second-order spatiotemporal accuracy.

\subsection{Three-phase Stefan problem: melting and boiling of a solid PCM} \label{sec_3pS_ana_num}

\begin{figure}[]
\label{result-threePhase}
    \begin{center}
        \subfigure[]{\includegraphics[scale=0.18]{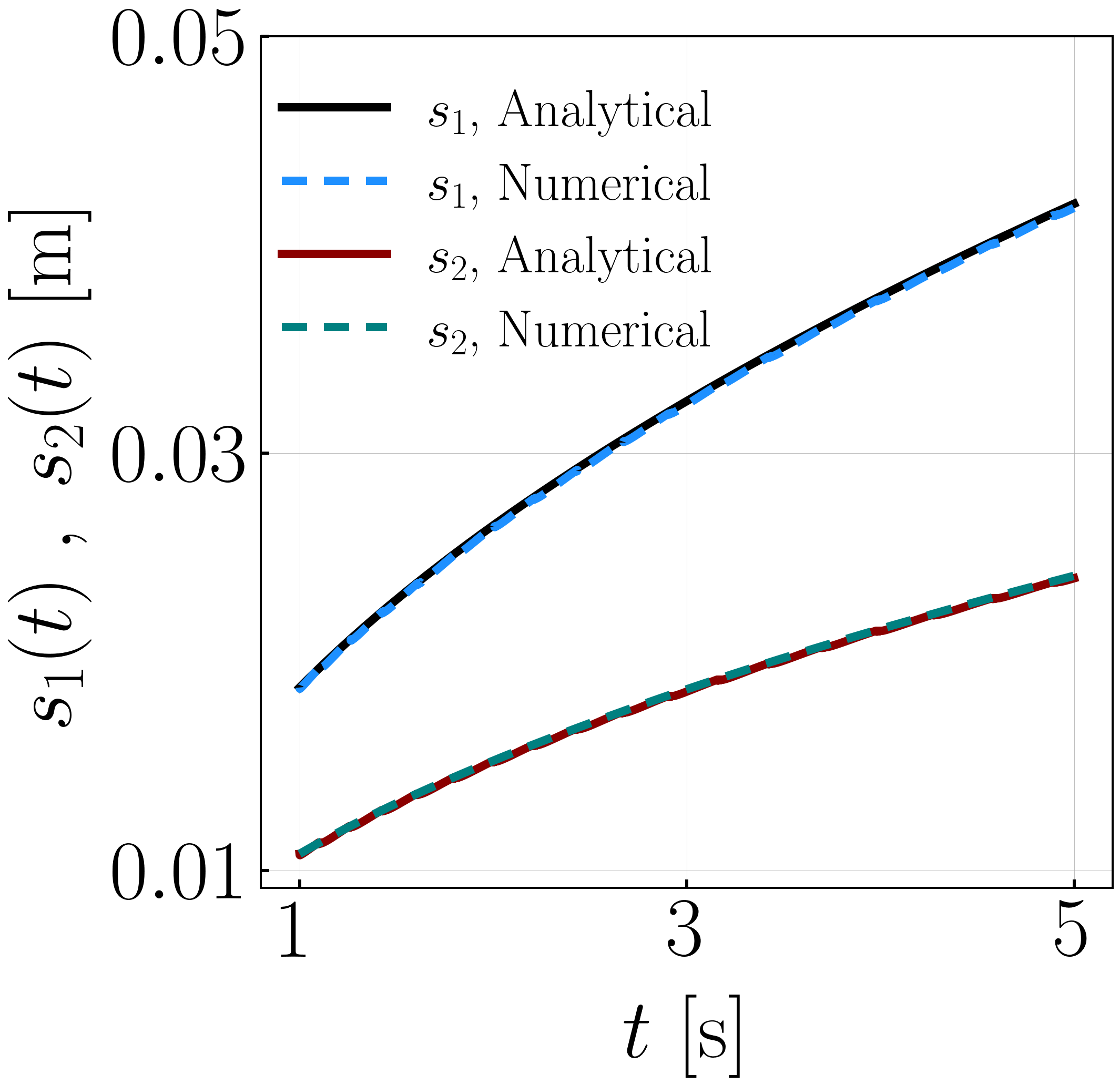}
        \label{fig_3p_A}
        }
        \hspace*{1.7cm}
        \subfigure[]{\includegraphics[scale=0.18]{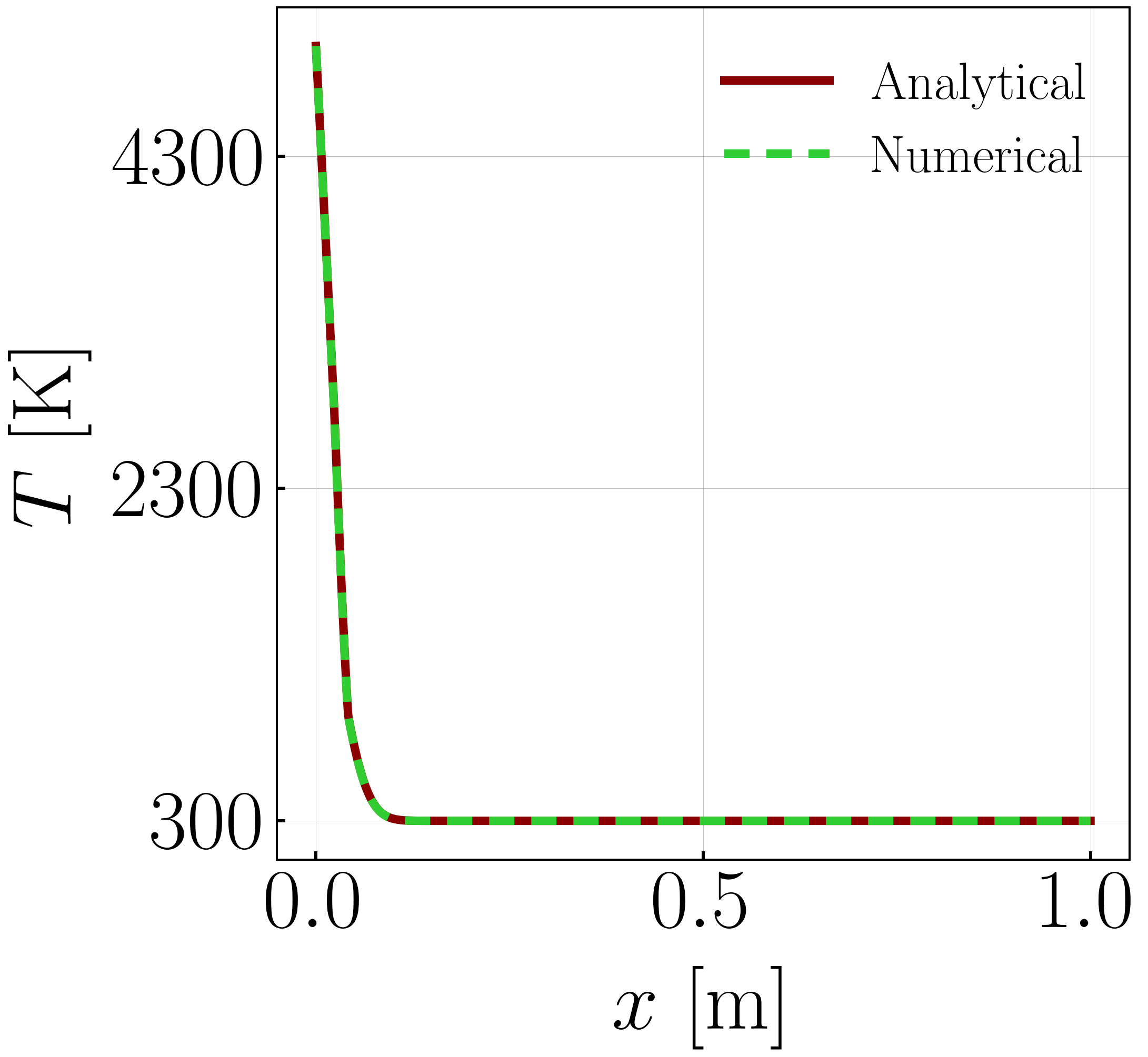}
        \label{fig_3p_B}
        }

        \subfigure[]{\includegraphics[scale=0.18]{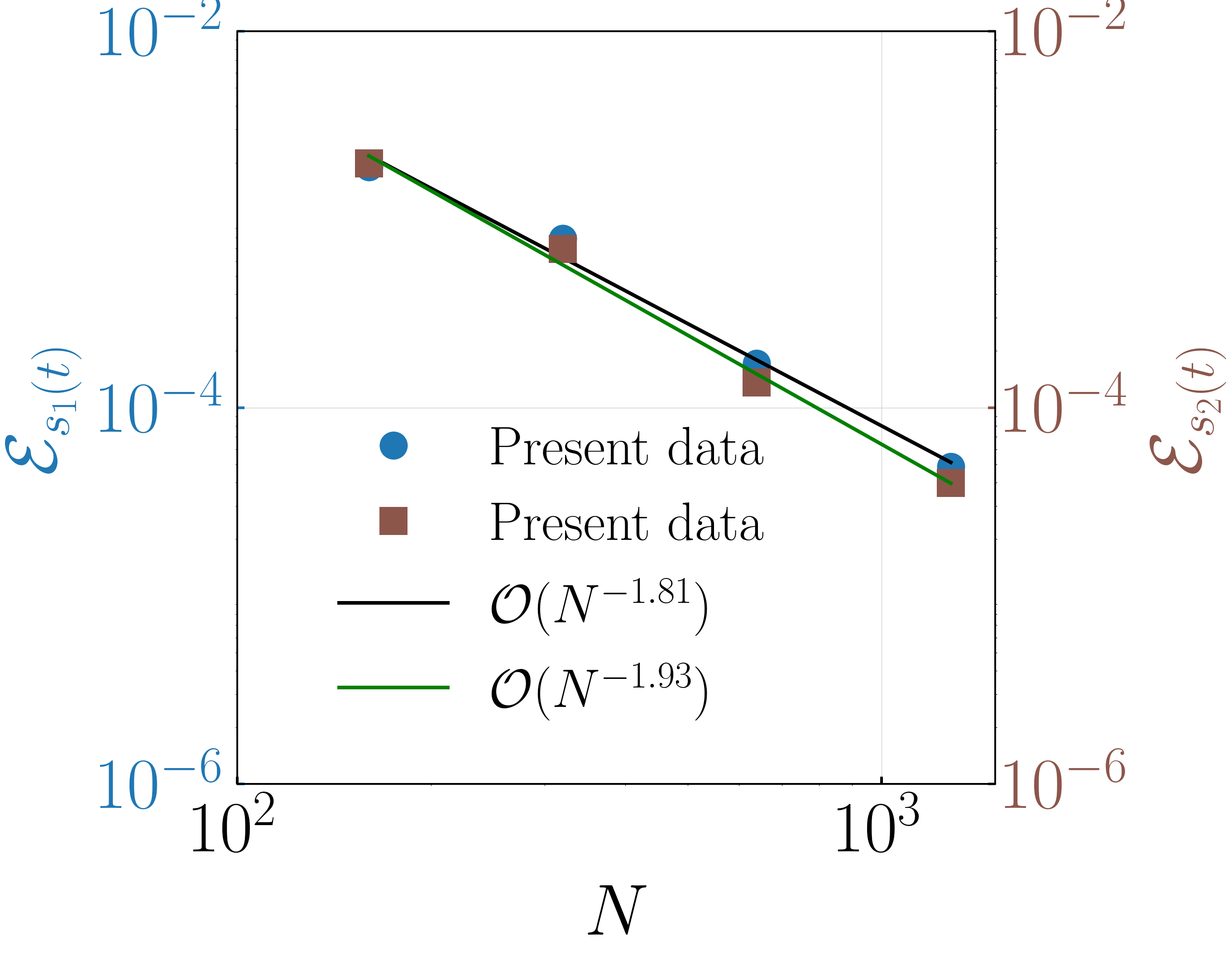}
        \label{fig_3p_C}
        }
         \hspace*{0.5cm}
        \subfigure[]{\includegraphics[scale=0.18]{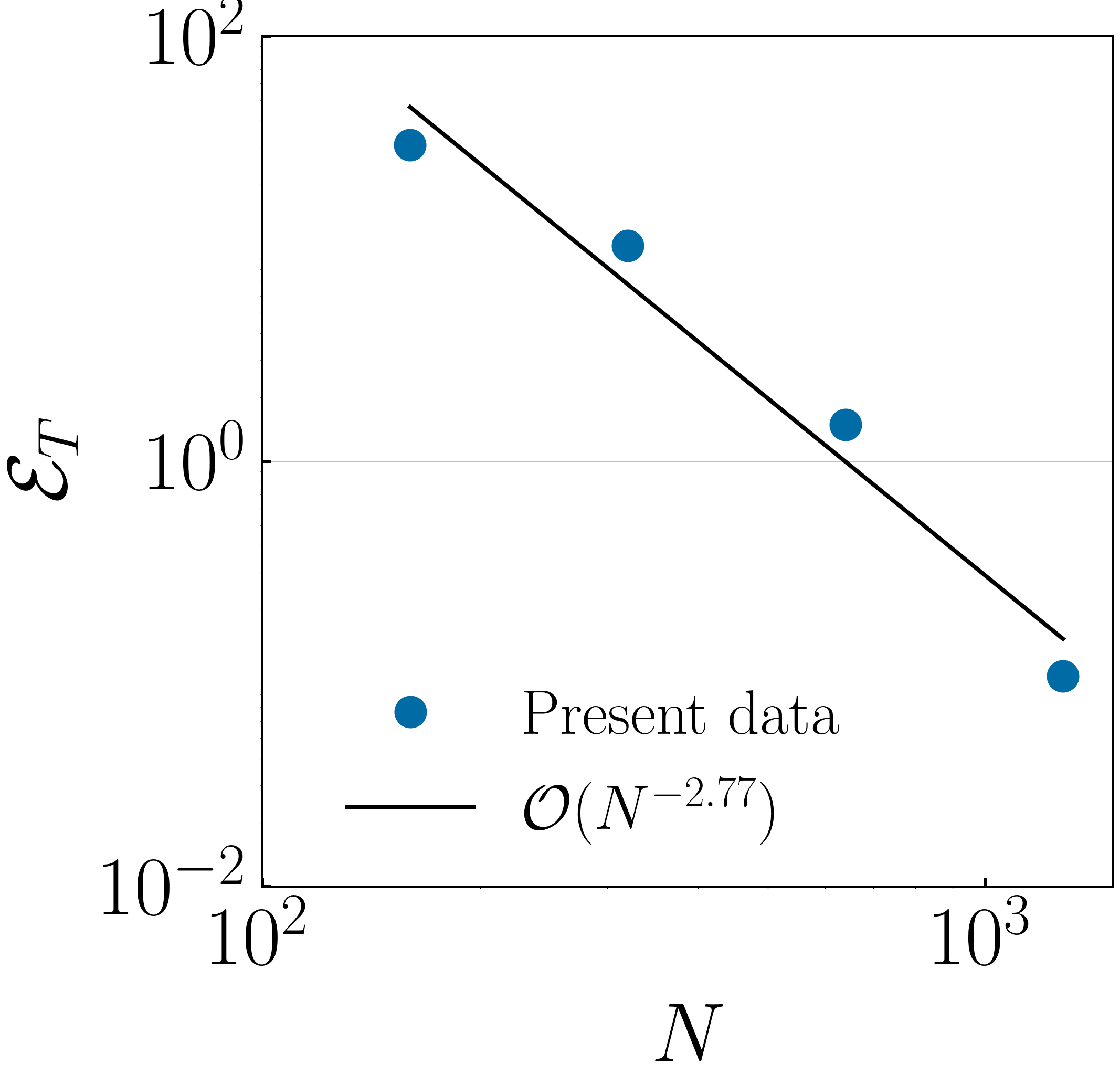}
        \label{fig_3p_D}
        }
    \end{center}
        \caption{Comparison of numerical results against the analytical solutions for the three-phase Stefan problem for \subref{fig_3p_A} vapor-liquid and liquid-solid  interface positions and \subref{fig_3p_B} temperature distribution across the domain. Convergence rate of the numerical error for \subref{fig_3p_C} interface position and \subref{fig_3p_D} temperature distribution as a function of grid size $N$.}
\end{figure}

The problem setup remains the same as discussed for the two-phase Stefan problem in Sec.~\ref{sec_2pS_ana_num}. The only difference is the imposed temperature at the left boundary, which is set to $T_\infty = 5000$ K. Because the imposed temperature is above the PCM's vaporization temperature ($T_v = 2767$ K), the initially solid PCM melts and boils simultaneously at $t = 0^{+}$. Due to the realistic thermophysical properties of the three phases, the melt front moves faster than the boiling front. 

To solve the three-phase Stefan problem numerically, we initialize the interface position and temperature in the domain based on the analytical solution at $t = 1$ s. The computational domain is discretized into $N = 1280$ cells. The time step size is taken to be $\Delta t = c \frac{\Delta x}{\text{max}\left(u_1^*,u_2^*\right)}$, in which $c = 0.0001$ is used as the CFL number. The interface velocities $u_1^*$ and $u_2^*$ are evaluated at $t = 1$ s, where they are at their maximum. The numerical simulation is run from $t = 1$ to $t = 5$ s. Figs.~\ref{fig_3p_A} and~\ref{fig_3p_B} compare the analytical and numerical solutions for the interface position and temperature in the domain, respectively. Both agree well.

Lastly, to determine the order of accuracy of the numerical method to simulate two evolving interfaces simultaneously, we conducted simulations on four grids of size: $N = \{ 160,320,640,1280\}$. The CFL number is fixed at $c = 0.0001$ for all grids. On each grid the simulation is run from $t = 6 $ to $t = 8$ s. Figs.~\ref{fig_3p_C} and~\ref{fig_3p_D} plot the actual error data as filled markers and best fit lines to determine the convergence rate of the error as a function of grid size $N$. It can be observed that the method converges with second-order spatiotemporal accuracy even for the three-phase Stefan problem with MSNBC. 

\section{Conclusions and discussion}

Based on the similarity transformation of the heat equation, we derive analytical solutions to the three-phase Stefan problem in this work. Solving the three-phase Stefan problem analytically requires solving two coupled transcendental equations. The transcendental equations are time-independent if the kinetic energy terms are omitted from the Stefan conditions. As part of this study, we also presented a sharp interface method for modeling the one-dimensional three-phase Stefan problem with melting, solidification, boiling, and condensation (MSNBC). To verify the scheme's second-order spatiotemporal accuracy, numerical solutions were compared to analytical ones. The main limitation of sharp-interface methods is that they require all phases within the domain to exist at the same time. With this approach, it is not trivial or perhaps even impossible to handle the sudden appearance or disappearance of a phase. We are consequently working towards extending the low Mach enthalpy method (which is a diffuse interface approach) to handle simultaneous MSNBC occurrences as well as sudden phase appearance or disappearance during metal manufacturing and welding processes. Nonetheless, the numerical results presented in this work provide confidence in the correctness of the analytical solutions. To conclude, this work provides benchmarks for CFD algorithms (such as the low Mach enthalpy method) used to simulate simultaneous MSNBC events based on verified analytical solutions.     

\section*{Acknowledgements}
A.P.S.B~acknowledges support of NSF award CBET CAREER 2234387. N.T~acknowledges support of NSF award 2306329. A.P.S.B dedicates this work to his brother, Sachin Kalra.

\section*{Appendix}
\begin{appendix}
\REVIEW{
\section{Comparison of the two front speeds}
\label{sec_compare_interface_speed}
 The two Stefan conditions of Eqs.~\eqref{eqn_Stefan_conditions} allow us to compare the speeds of melt and boiling fronts for the phase change material considered in this work. From Table~\ref{tab_thermophys_properties}, it can be observed that the thermal conductivity $\kappa$ and the specific heat $C$ of all three phases are in the same order of magnitude. For the purposes of quick comparison we will take them equal, i.e., $\cpl \approx \cps \approx \cpv \approx C$ and $\kl \approx \ks \approx \kv \approx \kappa$. Referring to Fig.~\ref{fig_3pstefan_schematic}, the temperature gradients in the three phases (assuming equal phase lengths $\Delta L$) can be approximated  as
\begin{subequations} 
\begin{alignat}{2}
&\frac{\partial T^{\rm L}}{\partial x} \approx \frac{T_m - T_v}{\Delta L}, \\
&\frac{\partial T^{\rm S}}{\partial x} \approx \frac{T_o - T_m}{\Delta L}, \\    
&\frac{\partial T^{\rm V}}{\partial x} \approx \frac{T_v - T_\infty}{\Delta L}.
\end{alignat}
\end{subequations}
Thus, from Eqs.~\eqref{eqn_Stefan_VL} and~\eqref{eqn_Stefan_LS}, the boiling and melt fronts move with speeds:
\begin{subequations} 
\begin{alignat}{2}
&u_2^* && \approx \kappa \Bigg( \frac{T_m - 2 T_v + T_\infty}{\rho^{\rm V}L_e \Delta L} \Bigg), \\
&u_1^* && \approx u_2^{*}  + \kappa \Bigg( \frac{T_o -  2T_m + T_v}{\rho^{\rm L}L_m \Delta L} \Bigg) - \Bigg(\frac{\rho^{\rm V}}{\rho^{\rm L}} \Bigg)u_2^* \nonumber \\
&  && = u_2^{*}  +  \frac{\kappa}{\rhol \Delta L} \underbrace{\Bigg( \frac{T_o -  2T_m + T_v}{L_m }  -  \frac{T_m - 2 T_v + T_\infty}{L_e} \Bigg)}_{> 0}.
\end{alignat}
\end{subequations}
Substituting the values of thermophysical properties from Table~\ref{tab_thermophys_properties}, it can be seen that $u_1^* > u_2^*$. In other words, for the PCM considered in this work, the melt front moves faster than the boiling front.
}
\end{appendix}
 
\bibliography{Ref.bib}

\begin{thebibliography}{10}
\expandafter\ifx\csname url\endcsname\relax
  \def\url#1{\texttt{#1}}\fi
\expandafter\ifx\csname urlprefix\endcsname\relax\def\urlprefix{URL }\fi
\expandafter\ifx\csname href\endcsname\relax
  \def\href#1#2{#2} \def\path#1{#1}\fi

\bibitem{thirumalaisamy2023low}
R.~Thirumalaisamy, A.~P.~S. Bhalla, A low mach enthalpy method to model
  non-isothermal gas--liquid--solid flows with melting and solidification,
  International Journal of Multiphase Flow 169 (2023) 104605.

\bibitem{vuik1993some}
C.~Vuik, {Some historical notes about the Stefan problem}, Tech. rep., Delft
  University of Technology, Faculty of Technical Mathematics and Informatics
  (1993).

\bibitem{rubinvsteuin2000stefan}
L.~I. Rubinvsteuin, {The Stefan problem}, Vol.~8, American Mathematical Soc.,
  2000.

\bibitem{hahn2012heat}
D.~W. Hahn, M.~N. {\"O}zisik, Heat conduction, John Wiley \& Sons, 2012.

\bibitem{alexiades2018mathematical}
V.~Alexiades, A.~D. Solomon, Mathematical modeling of melting and freezing
  processes, Routledge, 2018.

\bibitem{huang2022consistent}
Z.~Huang, G.~Lin, A.~M. Ardekani, A consistent and conservative phase-field
  model for thermo-gas-liquid-solid flows including liquid-solid phase change,
  Journal of Computational Physics 449 (2022) 110795.

\bibitem{yan2018fully}
J.~Yan, W.~Yan, S.~Lin, G.~Wagner, {A fully coupled finite element formulation
  for liquid--solid--gas thermo-fluid flow with melting and solidification},
  Computer Methods in Applied Mechanics and Engineering 336 (2018) 444--470.

\bibitem{javierre2006comparison}
E.~Javierre, C.~Vuik, F.~Vermolen, S.~Van~der Zwaag, {A comparison of numerical
  models for one-dimensional Stefan problems}, Journal of Computational and
  Applied Mathematics 192~(2) (2006) 445--459.

\bibitem{gibou2007level}
F.~Gibou, L.~Chen, D.~Nguyen, S.~Banerjee, {A level set based sharp interface
  method for the multiphase incompressible Navier--Stokes equations with phase
  change}, Journal of Computational Physics 222~(2) (2007) 536--555.

\bibitem{khalloufi2020adaptive}
M.~Khalloufi, R.~Valette, E.~Hachem, Adaptive eulerian framework for boiling
  and evaporation, Journal of Computational Physics 401 (2020) 109030.

\bibitem{thirumalaisamy2025consistent}
R.~Thirumalaisamy, A.~P.~S. Bhalla, A consistent, volume preserving, and
  adaptive mesh refinement-based framework for modeling non-isothermal
  gas--liquid--solid flows with phase change, International Journal of
  Multiphase Flow 183 (2025) 105060.

\bibitem{yu2022quantitative}
T.~Yu, J.~Zhao, Quantitative simulation of selective laser melting of metals
  enabled by new high-fidelity multiphase, multiphysics computational tool,
  Computer Methods in Applied Mechanics and Engineering 399 (2022) 115422.

\bibitem{ai2017three}
Y.~Ai, P.~Jiang, X.~Shao, P.~Li, C.~Wang, A three-dimensional numerical
  simulation model for weld characteristics analysis in fiber laser keyhole
  welding, International Journal of Heat and Mass Transfer 108 (2017) 614--626.

\bibitem{ai2017prediction}
Y.~Ai, P.~Jiang, X.~Shao, P.~Li, C.~Wang, G.~Mi, S.~Geng, Y.~Liu, W.~Liu, The
  prediction of the whole weld in fiber laser keyhole welding based on
  numerical simulation, Applied Thermal Engineering 113 (2017) 980--993.

\bibitem{nangia2019robust}
N.~Nangia, B.~E. Griffith, N.~A. Patankar, A.~P.~S. Bhalla, A robust
  incompressible navier-stokes solver for high density ratio multiphase flows,
  Journal of Computational Physics 390 (2019) 548--594.

\bibitem{doble2007perry}
M.~Doble, Perry’s chemical engineers’ handbook, McGraw-Hil, New York, US
  (2007).

\bibitem{hatch1984aluminium}
J.~E. Hatch, Aluminium: Properties and physical metallurgy, by asm, Metals
  Park, OH 135 (1984).

\bibitem{desai1987thermodynamic}
P.~Desai, Thermodynamic properties of aluminum, International journal of
  thermophysics 8 (1987) 621--638.

\end{thebibliography}
\end{document}